\documentclass[12pt, draftclsnofoot, onecolumn]{IEEEtran}
\ifCLASSINFOpdf
   \usepackage[pdftex]{graphicx}
   \graphicspath{{../pdf/}{../jpeg/}}
   \DeclareGraphicsExtensions{.pdf,.jpeg,.png}
\else
   \usepackage[dvips]{graphicx}
   \graphicspath{{../eps/}}
   \DeclareGraphicsExtensions{.eps}
\fi
\ifCLASSOPTIONcompsoc
\usepackage[caption=false, font=normalsize, labelfont=sf, textfont=sf]{subfig}
\else
\usepackage[caption=false, font=footnotesize]{subfig}

\usepackage{amsmath}
\usepackage{esint}
\usepackage{amssymb}
\usepackage{breqn}
\usepackage{cite}
\usepackage{color}

\ifCLASSOPTIONcompsoc
  \usepackage[caption=false,font=normalsize,labelfont=sf,textfont=sf]{subfig}
\else
  \usepackage[caption=false,font=footnotesize]{subfig}
\fi
\newcounter{MYtempeqncnt}
\newtheorem{theorem}{Theorem}
\newtheorem{lemma}{Lemma}

\begin{document}
%
\title{Near-Field Modelling and Performance Analysis
for Extremely Large-Scale IRS Communications}
\author{Chao~Feng, Haiquan~Lu, 
Yong~Zeng, ~\IEEEmembership{Senior Member, IEEE}, 
Teng~Li, ~\IEEEmembership{Member, IEEE},
Shi~Jin,  ~\IEEEmembership{Senior Member, IEEE}, 
and Rui~Zhang, ~\IEEEmembership{Fellow, IEEE}
\thanks{Part of this work has been published 
at the 2021 IEEE ICCC Workshops, Xiamen, China,
in July 2021 \cite{feng2021wireless}.
\emph{(Corresponding author: Yong Zeng.)}}
\thanks{Chao Feng and Shi Jin 
are with the National Mobile Communications Research Laboratory, 
Southeast University, Nanjing 210096, China 
(e-mail: chao\_feng@seu.edu.cn; jinshi@seu.edu.cn).}
\thanks{Haiquan Lu and Yong Zeng are with the National Mobile Communications Research
Laboratory, Southeast University, Nanjing 210096, China, and also
with Purple Mountain Laboratories, Nanjing 211111, China 
(e-mail: haiquanlu@seu.edu.cn; yong\_zeng@seu.edu.cn).}
\thanks{Teng Li is with the State Key Laboratory of Millimeter Waves, 
Southeast University, Nanjing 210096, China, and also
with Purple Mountain Laboratories, Nanjing 211111, China 
(e-mail: liteng@seu.edu.cn).}
\thanks{Rui Zhang is with The Chinese University of Hong Kong, Shenzhen, and
Shenzhen Research Institute of Big Data, Shenzhen, 518172, China (e-mail:
rzhang@cuhk.edu.cn). He is also with the Department of Electrical and
Computer Engineering, National University of Singapore, Singapore 117583
(e-mail: elezhang@nus.edu.sg).}
}
\maketitle

\begin{abstract}

Intelligent reflecting surface (IRS)
is an emerging technology for wireless communications,
thanks to its powerful capability to engineer the radio environment.
However, in practice,
this benefit is attainable 
only when the passive IRS is 
of sufficiently large size, 
for which the conventional uniform plane wave (UPW)-based far-field model 
may become invalid.
In this paper,
we pursue a near-field modelling and performance analysis
for wireless communications 
with extremely large-scale IRS (XL-IRS).
By taking into account the directional gain pattern 
of IRS's reflecting elements 
and the variations in signal amplitude 
across them,
we derive both the lower- and upper-bounds 
of the resulting signal-to-noise ratio (SNR)  
for the generic
uniform planar array (UPA)-based XL-IRS.
Our results reveal that, 
instead of scaling quadratically and unboundedly 
with the number of reflecting elements $M$
as in the conventional UPW-based model,
the SNR under the new non-uniform spherical wave (NUSW)-based model
increases with $M$ with a diminishing return
and eventually converges to a certain limit.
To gain more insights,
we further study the special case of 
uniform linear array (ULA)-based XL-IRS, 
for which a closed-form SNR expression 
in terms of the IRS size 
and locations of the base station (BS) and the user is derived.
Our result shows that 
the SNR is mainly determined by
the two geometric angles formed by the BS/user locations
with the IRS, as well as the dimension of the IRS.
Numerical results validate
our analysis and demonstrate the necessity  of proper near-field 
modelling for wireless communications aided by XL-IRS.

\end{abstract}


%
\IEEEpeerreviewmaketitle

\section{Introduction}
With the extensive deployment 
of the fifth-generation (5G) mobile communication networks, 
researchers from both industry and academia 
have started the investigation of beyond 5G (B5G) 
and sixth-generation (6G) wireless networks
\cite{latva2020key,you2021towards,saad2019vision,zeng2021toward}. 
To enhance the key performance metrics 
by orders-of-magnitude 
(e.g., data rate, latency, and connectivity density),   
several promising technologies have emerged,
such as extremely large-scale multiple-input multiple-output (XL-MIMO)
\cite{bjornson2019massive,de2020non,han2020channel,lu2020how,lu2021communicating,hu2018beyond_1,cui2022channel}, 
TeraHertz communication
\cite{rappaport2013millimeter,elayan2018terahertz},
and intelligent reflecting surface (IRS)
\cite{wu2021intelligentTutorial,wu2019intelligent,huang2019reconfigurable,wu2019towards,Tang2021WirelessCW,di2020smart,lu2021aerial}. 
In particular, IRS is a promising technology to 
achieve cost-effective and energy-efficient wireless communication 
by proactively manipulating the radio propagation environment
\cite{di2020smart,lu2021aerial,wu2019towards}. 
Compared with conventional relays, 
IRS-aided communication
dispenses with costly radio frequency (RF) chains and operates 
in a full-duplex mode, 
which is thus free of self-interference and noise amplification. 
However,
due to the double path-loss attenuation 
for signals reflected by IRS, 
to practically reap the promising performance gain 
of IRS-aided communications, 
the physical/electrical size of IRS 
needs to be sufficiently large\cite{bjornson2019demystifying,feng2021wireless}, 
leading to communication scenarios with
extremely large-scale IRS (XL-IRS). 
Note that the appealing features of lightweight and conformal geometry 
render it possible to deploy XL-IRSs in practice,  
such as on the facades of buildings, indoor walls and ceilings, etc. 

Most existing literatures on IRS-aided communications have focused on 
the conventional far-field uniform plane wave (UPW) assumption 
for ease of channel modelling and performance analysis, 
where all of IRS's reflecting elements share the identical angles
of arrival/departure (AoA/AoD) for the same channel path
\cite{wu2021intelligentTutorial,wu2019intelligent,wu2019towards,lu2021aerial}. 
Note that the typical criterion
to separate the far- and near-field propagation region 
is \emph{Rayleigh distance},
i.e., $r \geq \frac{2D^2}{\lambda}$,
where $r$ is the link distance,
$D$ and $\lambda$ denote the array physical dimension
and signal wavelength, respectively
\cite{selvan2017fraunhofer,yang2021communication,balanis2012advanced,balanis2015antenna,lu2021communicating}. 
As the aperture of IRS significantly increases, 
the transmitter and/or the receiver 
may no longer be located in the far-field region of XL-IRSs, 
and the conventional UPW-based signal model may become invalid\cite{feng2021wireless}.
Instead, 
the near-field signal characteristics 
with the more generic 
non-uniform spherical wave (NUSW) propagation
\cite{zhou2015spherical,friedlander2019localization}
should be considered to 
accurately model the variations in received signal amplitude, phase, 
and AoA/AoD across different reflecting elements of the IRS.
There have been several attempts
on the mathematical modelling and performance analysis 
for active antenna array communications 
under the near-field NUSW model
\cite{lu2021communicating,lu2021near}. 
For instance, 
in \cite{lu2021communicating},
by considering the variations 
in signal amplitude, phase and projected aperture
over different array elements, 
the authors derived
a closed-form expression for the received signal-to-noise ratio (SNR)
for XL-MIMO communications.
Additionally,
the multi-user XL-MIMO communication
was investigated in \cite{lu2021near}
based on three typical beamforming schemes, i.e.,
the maximal-ratio combining (MRC), zero-forcing (ZF), 
and minimum mean-square error (MMSE) beamforming.
Besides,
mathematical model considering the near-field signal characteristics
has been studied in the passive array/surface-based communications
\cite{dardari2020communicating,bjornson2020power,zheng2022simultaneous,bjornson2021primer}.
For example, in \cite{dardari2020communicating}, 
the impacts brought by distance variations in the near-field model were analyzed
for the continuous surface,
and accurate expressions for the achievable spatial degrees-of-freedom (DoFs)
of large intelligent surfaces (LIS)-based communications
were derived.
In \cite{bjornson2020power}, 
the power scaling laws and near-field behaviors of large-scale IRS 
were investigated under the two-dimensional (2D) modelling
without considering the elevation AoA/AoD.
In \cite{zheng2022simultaneous},
the asymptotic performance 
of IRS-aided simultaneous transmit diversity and passive beamforming, 
with the growing number of reflecting elements,
was derived in closed-form 
by considering the difference in signal amplitude and projected aperture 
over different reflecting elements.
Moreover, 
a new distance-based metric
was provided  
for accurate characterization 
of the antenna array gain 
in \cite{bjornson2021primer}.

Besides the NUSW-based near-field modelling, 
another important aspect of analyzing the performance 
of XL-IRS-aided communications is on 
accurately modelling the directional gain pattern of 
each individual reflecting element.
Most existing works have modelled each element's reflection pattern isotropically,
which, however, may lead to impractical results that
violate the law of power conservation
as the IRS size becomes large,
i.e., the total received power even exceeds the transmitted power.
To tackle this issue, 
directional gain pattern
has been widely considered in antenna theory
\cite{Tang2021WirelessCW,tang2022path,wang2021received,dardari2021nlos}. 
Specifically, 
in \cite{Tang2021WirelessCW} and \cite{tang2022path},
the impact of IRS reflecting element's directivity
was considered in the path loss modelling,
which was explicitly
expressed as an angle-dependent loss factor
and validated by experimental results.
In \cite{wang2021received}, 
a received power model for an IRS-aided wireless
communication system was proposed.
It was shown that 
the received power 
depends on the effective aperture of individual reflecting element,
and then
the radar cross section (RCS)
was adopted to characterize each element's directivity.
Besides,
by following the directional radiation gain pattern
commonly used in the active arrays\cite{balanis2015antenna}, 
the modelling with directional reflection gain pattern 
was further considered for IRS-aided communications
in \cite{dardari2021nlos} and \cite{ellingson2021path}.
However, 
the aforementioned works 
do not provide a rigorous near-field performance analysis, 
such as power scaling laws and asymptotic SNR expressions 
with respect to the number of IRS reflecting elements.

To fill the above gaps,
we study in this paper 
the three-dimensional (3D) near-field modelling and performance analysis 
for wireless communications aided by XL-IRS. 
The main contributions of this paper  
are summarized as follows.

\begin{itemize}
  \item Firstly, 
  by taking into account the directional gain pattern 
  of IRS's reflecting elements 
  and the variations in received signal amplitude and AoA/AoD
  across them,
  a generic near-field modelling is developed 
  for XL-IRS-aided communications
  based on the \emph{Friis Transmission Equation}\cite{friis1946note}.
  Based on the developed model, 
  tight lower- and
  upper-bounds of the received SNR are derived for the
  generic uniform planar array (UPA)-based XL-IRS.
  By further analysis,
  it is revealed that
  instead of scaling quadratically and unboundedly with the number
  of reflecting elements $M$
  as in the conventional UPW-based far-field model
  \cite{wu2019intelligent,lu2021aerial},
  the SNR under the new NUSW-based near-field model
  increases with $M$ with a diminishing return
  and eventually converges to a certain limit.
  \item Next,
  to gain further insights into our derived bounds,
  we also investigate the special case of 
  uniform linear array (ULA)-based XL-IRS, 
  for which a closed-form SNR expression 
  in terms of the IRS size 
  and locations of the base station (BS) and the user is derived.
  Similar to \cite{lu2020how},
  it is shown that 
  the SNR mainly depends on the two
  geometric angles formed by the BS/user locations
  with the IRS, as well as the dimension of the IRS.
  \item Lastly, 
  the developed model is extended to
  the multiple-input single-output (MISO) setup, 
  where the BS is equipped with multiple antennas.
  To comply with the typical deployment strategy where 
  the IRS is deployed
  closer to either the BS or the user 
  \cite{wu2021intelligentTutorial,lu2021aerial}, 
  we consider the scenario where
  the user is located in the near-field region of the IRS 
  while the BS is in the far-field region of it. 
  With the optimal single-user 
  maximum ratio combining/transmission (MRC/MRT) beamforming,
  an integral form of the received SNR
  is derived.
  Furthermore,
  we derive tight lower- and upper-bounds of the received SNR 
  in closed-form expressions 
  under a special gain pattern.
  The results reveal that 
  for the MISO scenario,
  there still exists a notable performance gap 
  between the conventional
  UPW model and the new near-field model as the IRS size becomes large.

\end{itemize}

The rest of this paper is organized as follows.
Section \uppercase\expandafter{\romannumeral2} 
introduces the near-field modelling 
for XL-IRS-aided communications. 
In Section \uppercase\expandafter{\romannumeral3}, 
tight lower- and upper-bounds of the SNR expression
are derived for the UPA-based XL-IRS,
and the special case of ULA-based XL-IRS is also considered.
We then extend our developed model to the MISO setup
with more insights given in Section \uppercase\expandafter{\romannumeral4}. 
Numerical results are provided 
in Section \uppercase\expandafter{\romannumeral5}. 
Finally, 
we conclude our paper in Section \uppercase\expandafter{\romannumeral6}.

\section{System Model}
\begin{figure}[ht]
  \centering
  \includegraphics[width=4.0in]{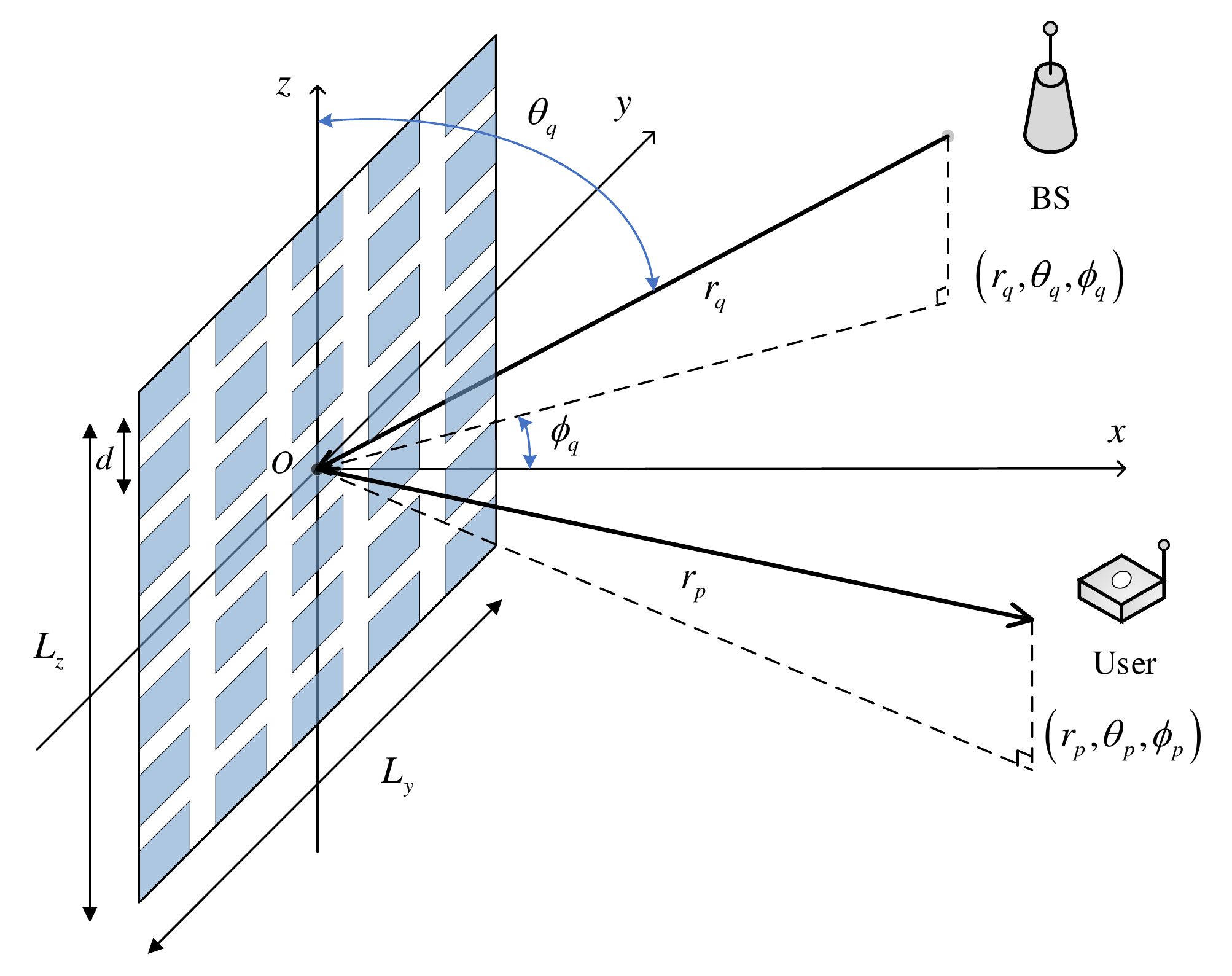}
  \caption{Wireless communication assisted by XL-IRS.}
  \label{XL-IRS-assisted MIMO communication_siso-1}
\end{figure}

As shown in Fig. \ref{XL-IRS-assisted MIMO communication_siso-1},
we consider a wireless communication system,
where an XL-IRS is deployed to assist in the communication 
between the BS and the user.
For ease of exposition,
we assume that the user is equipped with one antenna,
and the IRS is of UPA architecture.
The separation between adjacent elements
is denoted by $d \leq \frac{\lambda}2$,
where $\lambda$ denotes the signal wavelength.
It is assumed that 
the XL-IRS is located on the $y$-$z$ plane and centered at the origin,
and we have the total number of IRS reflecting elements as $M=M_y M_z \gg 1$
with $M_y$ and $M_z$ denoting 
the number of reflecting elements 
along the $y$- and $z$-axis, respectively.
Based on the above notations, the entire IRS size can be expressed as
$L_y \times L_z$, where $L_y \simeq M_y d$
and $L_z \simeq M_z d$.


For notational convenience,
we assume that 
both $M_y$ and $M_z$
are odd numbers.
Without loss of generality,
a Cartesian coordinate system is established 
so that the center of the XL-IRS coincides with the origin.
The central location of 
the $\left(m_y,m_z\right)$-th reflecting element is
denoted as $\mathbf{w}_{m_y,m_z}=\left[0,m_y d,m_z d\right]^T$,
where $m_y=0,\pm 1,\cdots,\pm \left(M_y-1\right)/2$,
$m_z=0,\pm 1,\cdots,\pm \left(M_z-1\right)/2$.
Denote the distance between the BS and the center of the XL-IRS as $r_q$,
and the location of the BS
is then denoted as 
$\mathbf{q}=\left[r_q \Psi_q,r_q \Phi_q,r_q \Theta_q\right]^T$,
with $\Psi_q \triangleq \sin\theta_q \cos\phi_q$,
$\Phi_q \triangleq \sin\theta_q \sin\phi_q$,
and $\Theta_q \triangleq \cos\theta_q$,
where $\theta_q \in \left[0,\pi\right]$
and $\phi_q \in \left[-\frac{\pi}2,\frac{\pi}2\right]$
denote the zenith and azimuth angles of the BS, respectively.
Thus, the distance between the BS 
and the $(m_y,m_z)$-th reflecting element can be expressed as 
\begin{equation}
  \label{eq_distance_BS_IRS}
  \begin{aligned}
    r_{q,m_y,m_z}
    =\left\|\mathbf{q}-\mathbf{w}_{m_y,m_z}\right\|
    =r_q\sqrt{1-2m_y\varepsilon_q\Phi_q-2m_z\varepsilon_q\Theta_q+\left(m_y^2+m_z^2\right)\varepsilon_q^2},
  \end{aligned}
\end{equation}
where $\varepsilon_q \triangleq \frac{d}{r_q}$.
Note that in practice,
we have $\varepsilon_q \ll 1$
\cite{lu2021communicating,feng2021wireless}.




Similarly,
denote the location of the user as
$\mathbf{p}=\left[r_p \Psi_p,r_p \Phi_p,r_p \Theta_p\right]^T$,
with $\Psi_p \triangleq \sin\theta_p \cos\phi_p$,
$\Phi_p \triangleq \sin\theta_p \sin\phi_p$,
and $\Theta_p \triangleq \cos\theta_p$,
where $r_p$ is the distance 
between the XL-IRS center and the user,
and $\theta_p \in \left[0,\pi\right]$
and $\phi_p \in \left[-\frac{\pi}2,\frac{\pi}2\right]$
denote the zenith and azimuth angles of the user, respectively.
As a result,
the distance between the $\left(m_y,m_z\right)$-th reflecting element 
and the user can be expressed as
\begin{equation}
  \label{eq_distance_rx}
  r_{p,m_{y},m_{z}}
  =r_p\sqrt{1-2m_y\varepsilon_p\Phi_p-2m_z\varepsilon_p\Theta_p+\left(m_y^2+m_z^2\right)\varepsilon_p^2},
\end{equation}
where $\varepsilon_p \triangleq \frac{d}{r_p} \ll 1$.

\begin{figure}[ht]
  \centering
  \includegraphics[width=3.0in]{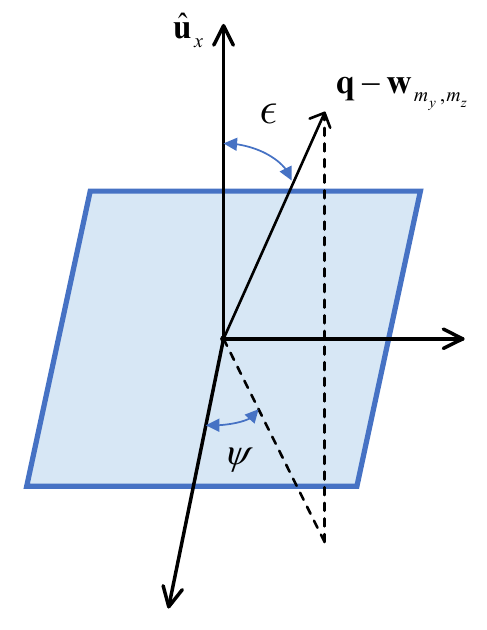}
  \caption{Illustration of the angles for the directional gain pattern for each IRS element.}
  \label{element_pattern_model}
\end{figure}

We assume that the direct link 
between the BS and the user is negligible.
In the preliminary version of this work
\cite{feng2021wireless},
we mainly pay attention to the variations in signal amplitude
across different reflecting elements, 
and the directional gain pattern of each element 
is essentially modelled by its projected aperture.
In this paper, we consider
a generic directional gain pattern
for each reflecting element given below \cite{balanis2015antenna}
\begin{equation}
  \label{eq_element_pattern}
  G_e(\epsilon,\psi)=
  \begin{cases}
    \gamma\prime \cos^{2q\prime}(\epsilon),
    &\epsilon \in \left[\left.0,\frac{\pi}2\right)\right.,
    \psi \in \left[0,2\pi\right]\\
    0,& \text{otherwise}
  \end{cases},
\end{equation}
where $\epsilon$ and $\psi$ 
are the elevation and azimuth angles
as shown in Fig. \ref{element_pattern_model},
$q\prime$ determines the directivity of the element,
and $\gamma\prime$ is the maximum gain 
in the boresight direction ($\epsilon=0$),
whose value depends on $q\prime$. 
According to the law of power conservation, 
we have \cite{balanis2015antenna}
\begin{equation}
  \label{eq_power_conservation}
  \varoiint_{S} G_e(\epsilon,\psi) \mathrm{d} \Omega
  =\int_0^{2\pi} \mathrm{d} \psi \int_0^{\frac{\pi}2} \gamma\prime \cos^{2q\prime}(\epsilon) \sin(\epsilon) \mathrm{d}\epsilon
  =4\pi,
\end{equation}
where $S$ is the surface of a semisphere
and $\Omega$ is the solid angle.
Therefore, 
the modelling parameters $\gamma\prime$ 
and $q\prime$ in \eqref{eq_element_pattern}
should follow the relationship
\cite{balanis2015antenna}
\begin{equation}
  \label{eq_pattern_relation}
  \gamma\prime=2(2q\prime+1).
\end{equation}

Furthermore, 
since the effective aperture of any antenna element
is proportional to its gain\cite{balanis2015antenna}, 
the maximum effective aperture $\mu$ of such a reflecting element is
\begin{equation}
  \label{eq_maximum effective aperture}
  \mu \triangleq \frac{\lambda^2}{4\pi} \gamma\prime
  =\frac{\lambda^2}{2\pi} (2q\prime+1).
\end{equation}

Based on the \emph{Friis Transmission Equation},
the ratio of the intercepted power by each IRS element
and that transmitted by a source can be
expressed as \cite{friis1946note,balanis2015antenna}
\begin{equation}
  \label{eq_Friis Transmission Equation_simplified}
  P_r/P_t=\left(\frac{\lambda}{4\pi r}\right)^2 G_t G_e(\epsilon,\psi),
\end{equation}
where $r$ is the link distance 
between the transmit antenna and a given IRS element,
and $G_t$ is the gain of the transmit antenna.
By assuming $G_t=1$, it then follows 
from \eqref{eq_distance_BS_IRS} and \eqref{eq_Friis Transmission Equation_simplified} that 
the channel power gain between 
the BS and the $\left(m_y,m_z\right)$-th reflecting element 
can be modelled as 
\begin{equation}
  \label{eq_power_gain_BS_general}
  a_{m_y,m_z}
  (r_q,\theta_q,\phi_q)
  =\left(\frac{\lambda}{4\pi r_{q,m_y,m_z}}\right)^2 \gamma\prime
  \cos^{2q\prime}(\epsilon_q),
\end{equation}
where $\cos(\epsilon_q)$ is 
the projection between the signal propagation direction
and the normal vector of the element surface, 
given by 
\begin{equation}
  \label{eq_cos_direction_BS}
  \cos(\epsilon_q)=
  \frac{\left(\mathbf{q}-\mathbf{w}_{m_y,m_z}\right)^T
  \hat{\mathbf{u}}_x}
  {\left\|\mathbf{q}-\mathbf{w}_{m_y,m_z}\right\|},
\end{equation}
with $\hat{\mathbf{u}}_x$ denoting a unit vector 
along the $x$-axis, and regarded as
the normal vector of any IRS reflecting element. 
By substituting \eqref{eq_cos_direction_BS} into \eqref{eq_power_gain_BS_general},
the channel power gain can be further expressed as
\eqref{eq_power_gain_BS} shown at the top of the next page.
Likewise,
the channel power gain between 
the $\left(m_y,m_z\right)$-th IRS element 
and the user can be expressed as \eqref{eq_power_gain_RX} 
at the top of the next page.

\begin{figure}[ht]
  \centering
  \includegraphics[width=4.0in]{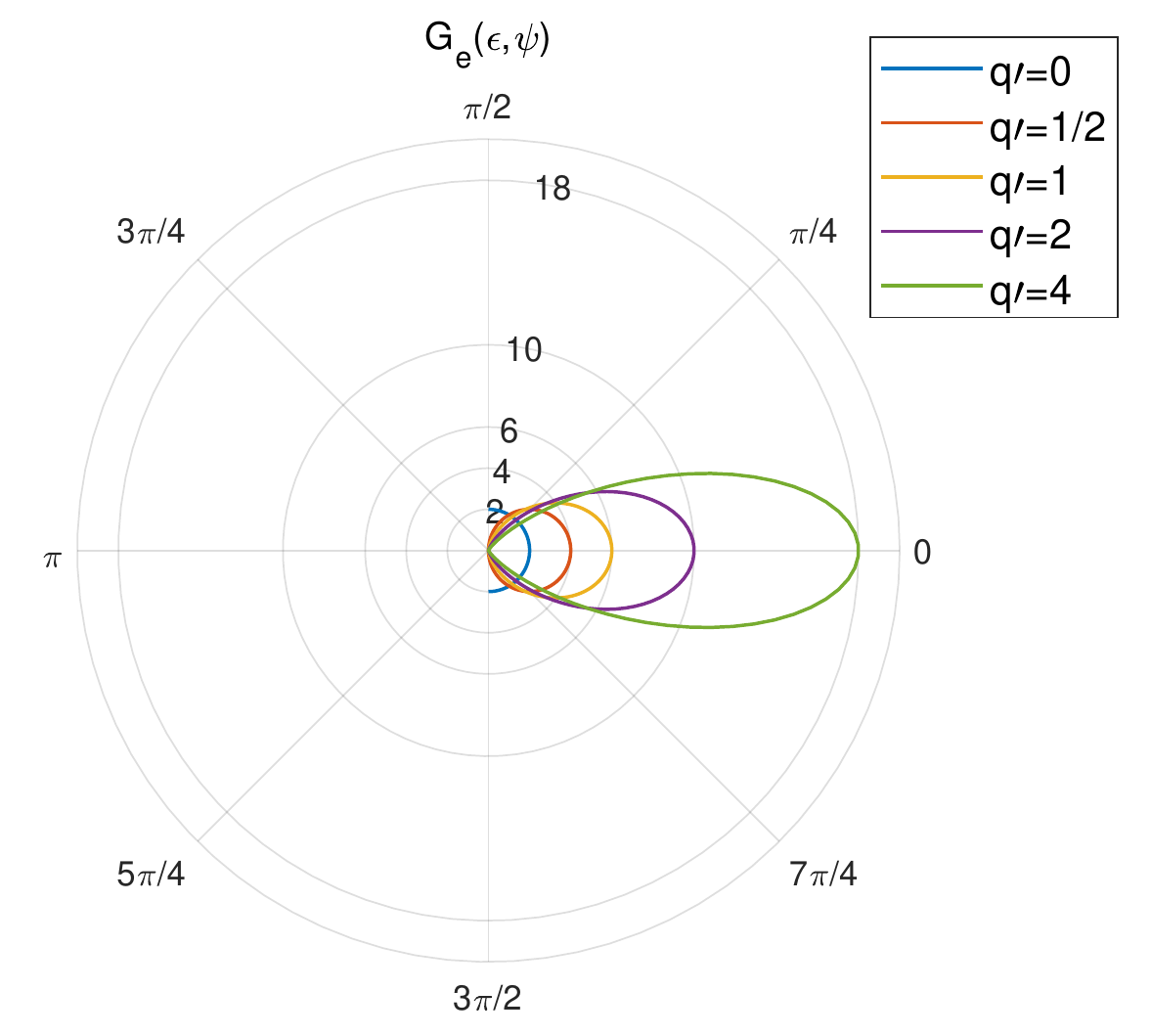}
  \caption{Directional gain pattern of IRS's reflecting elements for different parameters $q\prime$.}
  \label{Directional element pattern model}
\end{figure}

\begin{figure*}[!t]
  \normalsize
  \setcounter{MYtempeqncnt}{\value{equation}}
  \setcounter{equation}{\value{MYtempeqncnt}}

  \begin{equation}
    \label{eq_power_gain_BS}
  \begin{aligned}
  a_{m_y,m_z}
  (r_q,\theta_q,\phi_q)
  =\left(\frac{\lambda}{4\pi r_q}\right)^2 
  \frac{\gamma\prime \Psi_q^{2q\prime}}
  {\left[1-2m_y \varepsilon_q \Phi_q
  -2m_z \varepsilon_q \Theta_q
  +(m_y^2+m_z^2)\varepsilon_q^2\right]^{q\prime+1}}.
  \end{aligned}
  \end{equation}

  \setcounter{equation}{\value{MYtempeqncnt}+1}
  \hrulefill
  \vspace*{4pt}
  \end{figure*}

  \begin{figure*}[!t]
    \normalsize
    \setcounter{MYtempeqncnt}{\value{equation}}
    \setcounter{equation}{\value{MYtempeqncnt}}
  
    \begin{equation}
    \label{eq_power_gain_RX}
    b_{m_y,m_z}
    (r_p,\theta_p,\phi_p)
    =\left(\frac{\lambda}{4\pi r_p}\right)^2 
    \frac{\gamma\prime \Psi_p^{2q\prime}}
    {\left[1-2m_y \varepsilon_p \Phi_p
    -2m_z \varepsilon_p \Theta_p
    +(m_y^2+m_z^2)\varepsilon_p^2\right]^{q\prime+1}}.
    \end{equation}
  
    \setcounter{equation}{\value{MYtempeqncnt}+1}
    \hrulefill
    \vspace*{4pt}
  \end{figure*}

The channel vector between BS and the XL-IRS
is denoted as
$\mathbf{h}\in \mathbb{C}^{M\times 1}$,
whose entries are given by
\begin{equation}
  \label{eq_channel_vector_BS}
  h_{m_y,m_z}=\sqrt{a_{m_y,m_z}
  \left(r_q,\theta_q,\phi_q\right)}
  e^{-j\frac{2\pi}{\lambda}r_{q,m_y,m_z}},
  \forall m_y,m_z.
\end{equation}

Similarly,
the channel vector between the XL-IRS and the user
is denoted as
$\mathbf{g}\in \mathbb{C}^{M \times 1}$,
whose entries are given by
\begin{equation}
  \label{eq_channel_vector_RX}
  g_{m_y,m_z}=\sqrt{b_{m_y,m_z}
  \left(r_p,\theta_p,\phi_p\right)}
  e^{-j\frac{2\pi}{\lambda}r_{p,m_y,m_z}},
  \forall m_y,m_z.
\end{equation}

Further denote the phase shift introduced by 
the $(m_y,m_z)$-th IRS element
as $\theta_{m_y,m_z}$,
and $\mathbf{\Theta}\in \mathbb{C}^{M \times M}$ 
is a diagonal matrix, with the diagonal entries  
given by $e^{j\theta_{m_y,m_z}}$.
Then the received signal at the user 
can be obtained as
\begin{equation}
  \label{eq_received_signal}
  y=\mathbf{g}^T \mathbf{\Theta} \mathbf{h}
  \sqrt{P} s+n,
\end{equation}
where
$P$ and $s$ denote the transmit power 
and information-bearing symbol, respectively;
$n \sim \mathcal{CN}(0,\sigma^2)$
is the additive white Gaussian noise (AWGN)
at the user.

Thus, the SNR at the user can be formulated as
\begin{equation}
  \label{eq_received_snr}
  \gamma=\bar{P}\left|\mathbf{g}^T \mathbf{\Theta} \mathbf{h}\right|^2,
\end{equation}
where $\bar{P} \triangleq \frac{P}{\sigma^2}$.

\section{Performance Analysis}
In this section, 
we mainly focus on the performance analysis 
including the SNR's lower- and upper-bounds and its asymptotic analysis
as $M$ goes to infinity,
based on the generic XL-IRS-aided system model 
and the SNR expression in \eqref{eq_received_snr}.


\subsection{SNR Lower- and Upper-Bounds}

From \eqref{eq_received_snr}, 
the optimal phase shifting by the XL-IRS can be given by
\begin{equation}
  \label{eq_optimal_phase_shift}
  \theta_{m_y,m_z}=\frac{2\pi}{\lambda}r_{q,m_y,m_z}
  +\frac{2\pi}{\lambda}r_{p,m_y,m_z}.
\end{equation}

Therefore, the maximum SNR at the user can be expressed as 
\begin{equation}
  \label{eq_max_snr}
  \gamma=\bar{P}
  \left(\sum_{m_y}
  \sum_{m_z}
  \left|h_{m_y,m_z}\right|\left|g_{m_y,m_z}\right|\right)^2.
\end{equation}

By substituting 
\eqref{eq_power_gain_BS}, \eqref{eq_power_gain_RX}, 
\eqref{eq_channel_vector_BS} and \eqref{eq_channel_vector_RX} 
into \eqref{eq_max_snr},
we can obtain the resulting SNR as \eqref{eq_snr_summation}, 
shown at the top of the page.
Furthermore,
motivated by the similar methods in \cite{lu2021communicating},
the double summation in \eqref{eq_snr_summation}
can be approximately transformed into its corresponding double integral 
based on the fact that $\varepsilon_q\ll 1$ and $\varepsilon_p\ll 1$.
Thus, by following the above approximation, 
the received SNR can be rewritten in an integral form 
as \eqref{eq_snr_integral}, shown at the top of the page.

\begin{figure*}[!t]
  \normalsize
  \setcounter{MYtempeqncnt}{\value{equation}}
  \setcounter{equation}{\value{MYtempeqncnt}}

  \begin{equation}
  \begin{aligned}
  \gamma=\left(\frac{\lambda}{4\pi}\right)^4&
  \frac{\gamma\prime^2 \bar{P} \Psi_q^{2q\prime} \Psi_p^{2q\prime}}{r_q^2 r_p^2}
  \left|\sum_{m_y}\sum_{m_z} 
  \frac1{\left[1-2m_y\varepsilon_q\Phi_q-2m_z\varepsilon_q\Theta_q+(m_y^2+m_z^2)\varepsilon_q^2\right]^{(q\prime+1)/2} } \right.\\
  &\left.\times\frac1{\left[1-2m_y\varepsilon_p\Phi_p-2m_z\varepsilon_p\Theta_p+(m_y^2+m_z^2)\varepsilon_p^2\right]^{(q\prime+1)/2}} \right| ^2.
  \end{aligned}
  \label{eq_snr_summation}
  \end{equation}

  \setcounter{equation}{\value{MYtempeqncnt}+1}
  \hrulefill
  \vspace*{4pt}
\end{figure*}

\begin{figure*}[!t]
  \normalsize
  \setcounter{MYtempeqncnt}{\value{equation}}
  \setcounter{equation}{\value{MYtempeqncnt}}
  
  \begin{equation}
    \label{eq_snr_integral}
    \begin{aligned}
      &\gamma
      \simeq 
      \left(\frac{\lambda}{4\pi}\right)^4
      \frac{\gamma\prime^2 \bar{P} \Psi_q^{2q\prime} \Psi_p^{2q\prime}}{d^4 r_q^2 r_p^2}\\
      &\times
      \left|
      \int_{-\frac{L_z}2}^{\frac{L_z}2}
      \int_{-\frac{L_y}2}^{\frac{L_y}2}
      \frac{\mathrm{d} y \mathrm{d} z}{
      \left[1-\frac2{r_q} y\Phi_q
      -\frac2{r_q} z\Theta_q+\frac1{r_q^2}(y^2+z^2)\right]^{(q\prime+1)/2}
      \left[1-\frac2{r_p} y\Phi_p
      -\frac2{r_p} z\Theta_p
      +\frac1{r_p^2}(y^2+z^2)\right]^{(q\prime+1)/2} }
      \right|^2.
    \end{aligned}
  \end{equation}
  
  \setcounter{equation}{\value{MYtempeqncnt}+1}
  \hrulefill
  \vspace*{4pt}
\end{figure*}

Note that
obtaining a closed-form expression for the double integral 
in \eqref{eq_snr_integral} is challenging in general. 
In the following, 
we will first discuss some special cases for $q\prime$ in \eqref{eq_element_pattern}.
\begin{itemize}
  \item{(1)} Semi-isotropic pattern: $q\prime=0$ and $\gamma\prime=2$, so that
  \begin{equation}
    \label{eq_element_pattern_istropic}
    G_e(\epsilon,\psi)=
    \begin{cases}
      2,&\epsilon \in \left[\left.0,\frac{\pi}2\right)\right.,
      \psi \in \left[0,2\pi\right]\\
      0,& \text{otherwise}
    \end{cases}
  \end{equation}
  \item{(2)} Cosine pattern (based on projected aperture): 
  $q\prime=\frac12$ and $\gamma\prime=4$, so that
  \begin{equation}
    \label{eq_element_pattern_effective_aperture}
    G_e(\epsilon,\psi)=
    \begin{cases}
      4 \cos(\epsilon),
      &\epsilon \in \left[\left.0,\frac{\pi}2\right)\right.,
      \psi \in \left[0,2\pi\right]\\
      0,& \text{otherwise}
    \end{cases}
  \end{equation} 
  \item{(3)} Cosine-square pattern: $q\prime=1$ and $\gamma\prime=6$, so that
  \begin{equation}
    \label{eq_element_pattern_square}
    G_e(\epsilon,\psi)=
    \begin{cases}
      6 \cos^2(\epsilon),
      &\epsilon \in \left[\left.0,\frac{\pi}2\right)\right.,
      \psi \in \left[0,2\pi\right]\\
      0,& \text{otherwise}
    \end{cases}
  \end{equation} 
\end{itemize}

Fig. \ref{Directional element pattern model}
plots different directional gain patterns of IRS's reflecting elements
in \eqref{eq_element_pattern}
versus the angle $\epsilon$.
It can be observed that power will be radiated into just
one half of the space, i.e., $0 \leq \epsilon \leq \pi/2$.
Besides,
for a larger directivity parameter $q\prime$,
there exists a stronger beam 
with respect to the element's boresight direction ($\epsilon=0$).

\begin{figure}[ht]
  \centering
  \includegraphics[width=4.0in]{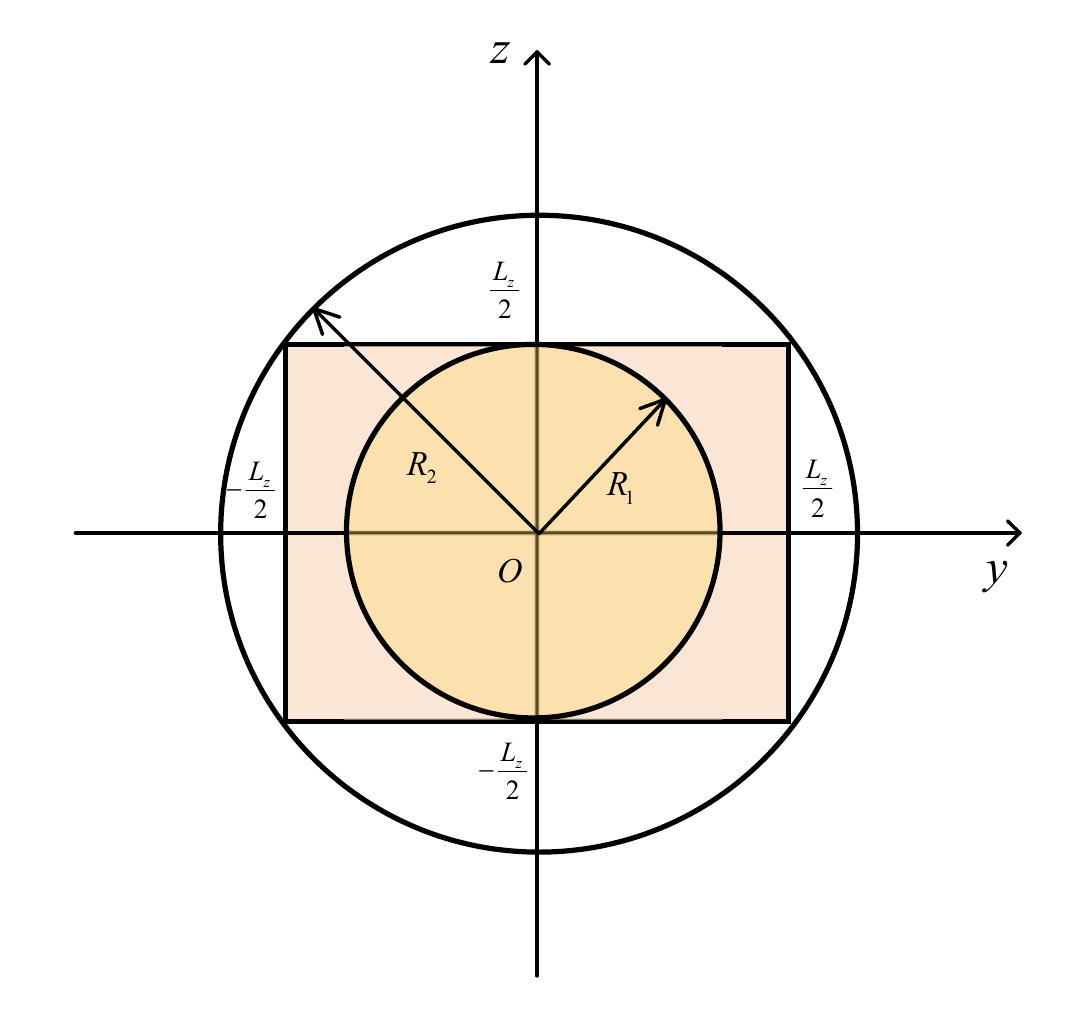}
  \caption{The inscribed and circumscribed disks of the rectangular region $L_y \times L_z$ occupied by the XL-IRS.}
  \label{inscribed and circumscribed disk}
\end{figure}

\begin{theorem}
  \label{theorem_bounds}
  For XL-IRS-aided communication,
  the resulting SNR in \eqref{eq_snr_integral}  
  is bounded by
  \begin{equation}
    \label{eq_siso_general_bounds}
    f(R_1,q\prime) \leq \gamma \leq f(R_2,q\prime),
  \end{equation}
  where the function $f(R,q\prime)$ is defined 
  as \eqref{eq_siso_general_bounds_function}
  shown at the top of the page;
  \begin{figure*}[!t]
    \normalsize
    \setcounter{MYtempeqncnt}{\value{equation}}
    \setcounter{equation}{\value{MYtempeqncnt}}
      
    \begin{equation}
      \label{eq_siso_general_bounds_function}
      \begin{aligned}
        &f(R,q\prime) \triangleq 
        \left(\frac{\lambda}{4\pi}\right)^4
        \frac{\gamma\prime^2 \bar{P} \Psi_q^{2q\prime} \Psi_p^{2q\prime}}
        {d^4 r_q^2 r_p^2}  \\
        &\times
        \left|\int_{0}^{2\pi} \mathrm{d} \zeta
        \int_{0}^{R} 
        \frac{r \mathrm{d} r}{
        \left[1-\frac{2r}{r_q} \Phi_q\cos\zeta
        -\frac{2r}{r_q} \Theta_q\sin\zeta
        +\frac{r^2}{r_q^2}\right]^{(q\prime+1)/2}
        \left[1-\frac{2r}{r_p} \Phi_p\cos\zeta 
        -\frac{2r}{r_p} \Theta_p\sin\zeta
        +\frac{r^2}{r_p^2}\right]^{(q\prime+1)/2} }
        \right|^2. 
      \end{aligned}
    \end{equation}
      
    \setcounter{equation}{\value{MYtempeqncnt}+1}
    \hrulefill
    \vspace*{4pt}
  \end{figure*}
  $R_1$ and $R_2$ denote the radii of the inscribed disk
  and circumscribed disk of the rectangular region
  $L_y \times L_z$ occupied by the XL-IRS 
  as shown in Fig. \ref{inscribed and circumscribed disk},
  given by
  \begin{equation}
    \label{eq_inscribed_with_circumscribed}
    \begin{aligned}
      &R_1=\frac12 \min\left\{L_y,L_z\right\},\\
      &R_2=\frac12 \sqrt{L_y^2+L_z^2}.
    \end{aligned}
  \end{equation}
    
\end{theorem}

\begin{IEEEproof}
  Note that the resulting SNR is expressed as a double integral form 
  over the region $L_y \times L_z$,
  based on the Cartesian coordinate system,
  and the integrand is meanwhile positive.
  Theorem \ref{theorem_bounds}
  can be verified by 
  scaling this rectangular integral region
  with its inscribed disk and circumscribed disk that have radii
  $R_1$ and $R_2$, respectively.
  It then follows from the variable transformations that
  lower- and upper-bounds of the SNR in \eqref{eq_siso_general_bounds} 
  can be obtained as an integral in polar coordinate.
\end{IEEEproof}

For convenience, we define a distance ratio
as $\rho \triangleq r_q/r_p$,
which represents the ratio of the distance
from the BS to the center of the XL-IRS
over that from the user to the XL-IRS center.
Without loss of generality, it can be assumed that 
$0 < \rho \leq 1$ due to symmetry.

\begin{lemma}
  \label{lemma_boresight}
  If the BS and the user 
  are both located along the boresight of the XL-IRS, 
  i.e., 
  near the $x$-axis
  with $\Phi_q,\Phi_p \ll \frac{r_q}{L_y}$ and 
  $\Theta_q,\Theta_p \ll \frac{r_q}{L_z}$,
  we have

  \begin{equation}
    \label{eq_snr_special_bounds_integral}
    \frac{\mu^2 \bar{P}}{4 d^4} G(R_1,q\prime) \leq \gamma 
    \leq \frac{\mu^2 \bar{P}}{4 d^4} G(R_2,q\prime),
  \end{equation}
  where $\mu$ is the maximum \emph{effective aperture} of each reflecting element
  defined in \eqref{eq_maximum effective aperture},
  and the function $G(R,q\prime)$ is defined as
  \begin{equation}
    \label{eq_G_integral}
    G(R,q\prime) \triangleq
    \left[ \rho
    \int_0^{\arctan \frac{R}{r_q}} 
    \frac{\cos^{2q\prime}\alpha \tan\alpha \mathrm{d}\alpha}{\left[\rho^2+(1-\rho^2)\cos^2 \alpha\right]^{(q\prime+1)/2}}
    \right]^2.
  \end{equation}
    
\end{lemma}

\begin{IEEEproof}
  Please refer to Appendix A.
\end{IEEEproof}

\begin{lemma}
  \label{bounds_peojected_aperture}
  For the special case 
  with the cosine gain pattern
  based on the projected aperture in \eqref{eq_element_pattern_effective_aperture}, 
  i.e., $q\prime=\frac12$, 
  under the same condition as Lemma \ref{lemma_boresight}, 
  lower- and upper-bounds of the SNR 
  can be further expressed as 
  \begin{equation}
    \label{eq_snr_special_bounds_1}
    \begin{cases}
      \frac{\mu^2 \bar{P}}{4 d^4} G\left(R_1,\frac12\right) \leq \gamma 
      \leq \frac{\mu^2 \bar{P}}{4 d^4} G\left(R_2,\frac12\right),& 0<\rho<1\\
      \\
      \gamma=\frac{\mu^2 \bar{P}}{\pi^2 d^4} \arctan^2 
      \frac{\left(\frac{L_y}{2 r_q}\right)\left(\frac{L_z}{2 r_q}\right)}
      {\sqrt{\left(\frac{L_y}{2 r_q}\right)^2+\left(\frac{L_z}{2 r_q}\right)^2+1}},& \rho=1
    \end{cases},
  \end{equation}
  where the function $G\left(R,\frac12\right)$ with $0 < \rho < 1$ can be 
  obtained in closed-form as
  \begin{equation}
    \label{eq_function_G}
    \begin{aligned}
      G
      &\left(R,\frac12\right)
      \triangleq 
      \frac{4\rho}{1-\rho^2}
      \Bigg[F\left(\frac12 \arctan \frac{\sqrt{1-\rho^2}}{\rho}\bigg|2\right)\\
      &\qquad-F\left(\frac12 \arctan\left(\frac{\sqrt{1-\rho^2}}{\rho}
      \cos\left(\arctan \frac{R}{r_q}\right)\right)\bigg|2\right)
      \Bigg]^2,
    \end{aligned}
  \end{equation}
  and $F(\vartheta|k)=\int_0^{\vartheta} \frac{1}{\sqrt{1-k\sin^2\beta}}\mathrm{d} \beta$
  is the incomplete Elliptic Integral of the First Kind \cite{luke1968approximations}.

\end{lemma}

\begin{IEEEproof}
  The proof follows the similar steps as Appendix A of \cite{feng2021wireless}, 
  which is omitted for brevity. 
\end{IEEEproof}

\begin{lemma}
  \label{bounds_square_pattern}
  For the special case with the cosine-square gain pattern  
  in \eqref{eq_element_pattern_square}, i.e., 
  $q\prime=1$, 
  under the same condition as Lemma \ref{lemma_boresight}, 
  lower- and upper- bounds of the SNR 
  can be further expressed as 
  \begin{equation}
    \label{eq_snr_special_bounds_2}
    \frac{\mu^2 \bar{P}}{4 d^4} G(R_1,1) \leq \gamma 
    \leq \frac{\mu^2 \bar{P}}{4 d^4} G(R_2,1),
  \end{equation}
  where the function $G(R,1)$ 
  is given in closed-form as \eqref{eq_function_G_2},
  shown at the top of the page.
\end{lemma}

\begin{figure*}[!t]
  \normalsize
  \setcounter{MYtempeqncnt}{\value{equation}}
  \setcounter{equation}{\value{MYtempeqncnt}}
  \begin{equation}
    \label{eq_function_G_2}
    \begin{aligned}
      G(R,1) 
      \triangleq
      \begin{cases}
        \frac{\rho^2}{4\left(1-\rho^2\right)^2}
        \Bigg[
        \ln \left[\rho^2+(1-\rho^2)\cos^2\left(\arctan \frac{R}{r_q}\right)\right]  
      \Bigg]^2, &0 < \rho < 1\\
      \\
      \frac14 \Bigg[\cos^2\left(\arctan \frac{R}{r_q}\right)-1
      \Bigg]^2,
      &\rho=1
      \end{cases}.
    \end{aligned}
  \end{equation}
  \setcounter{equation}{\value{MYtempeqncnt}+1}
  \hrulefill
  \vspace*{4pt}
\end{figure*}

\begin{IEEEproof}
  Please refer to Appendix B.
\end{IEEEproof}

\subsection{Asymptotic Analysis}

\begin{figure*}[!t]
  \normalsize
  \setcounter{MYtempeqncnt}{\value{equation}}
  \setcounter{equation}{\value{MYtempeqncnt}}
  \begin{equation}
    \label{eq_snr_special_asymptotic_integral_with relation}
    \begin{aligned}
      \lim_{L_y,L_z \rightarrow \infty} \gamma = 
        \frac{\lambda^4 \rho^2}{16 \pi^2 d^4}
        (2 q\prime+1)^2 \bar{P}
        \left[ 
        \int_0^{\pi/2} 
        \frac{\cos^{2q\prime}\alpha \tan\alpha \mathrm{d} \alpha}{\left[\rho^2+(1-\rho^2)\cos^2 \alpha\right]^{(q\prime+1)/2}}
        \right]^2.
    \end{aligned}
  \end{equation}
  \setcounter{equation}{\value{MYtempeqncnt}+1}
  \hrulefill
  \vspace*{4pt}
\end{figure*}

\begin{figure}[ht]
  \centering
  \includegraphics[width=4.0in]{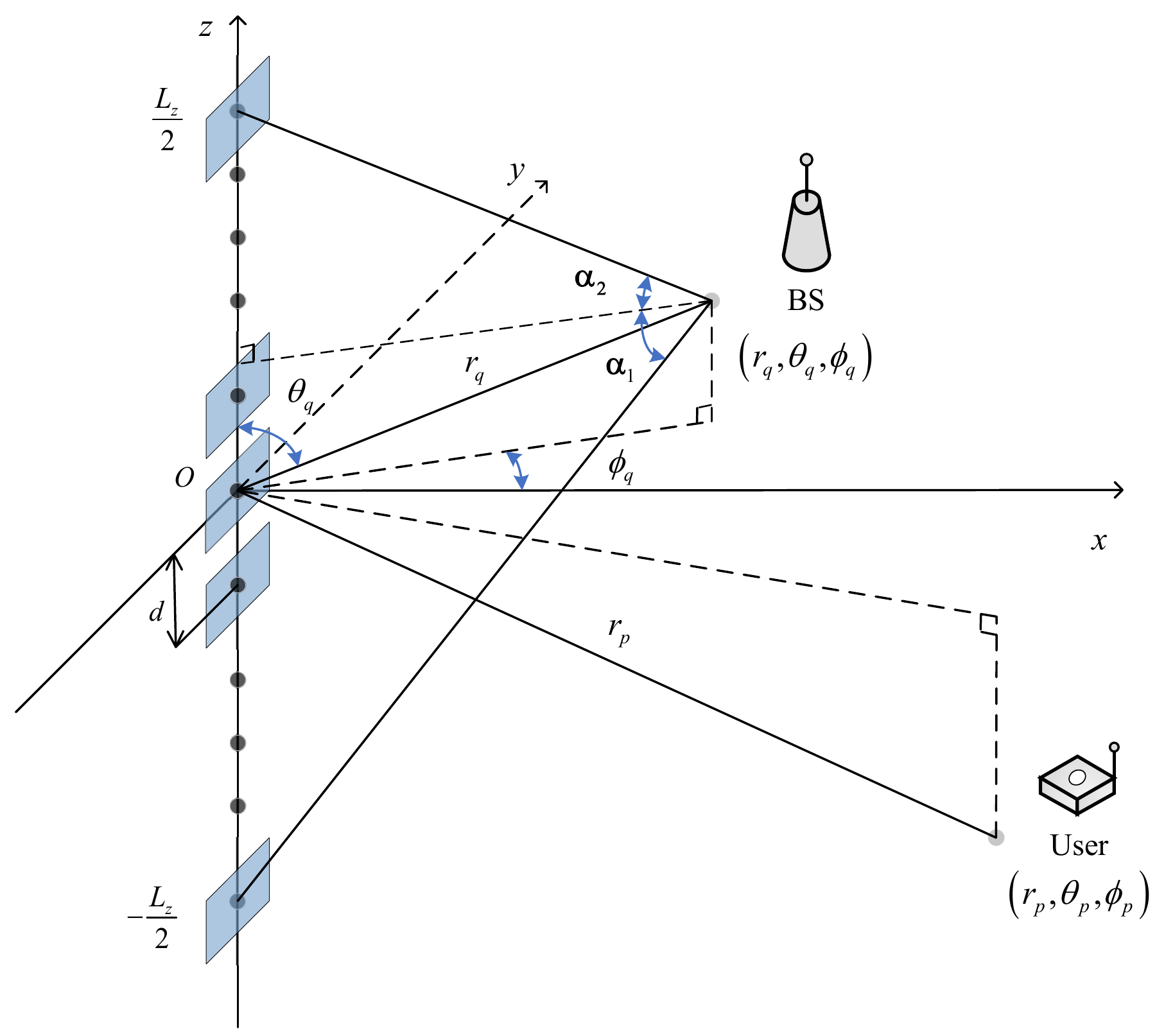}
  \caption{Wireless communication with ULA-based XL-IRS.}
  \label{IRS_ULA_model}
\end{figure}

\begin{lemma}
  \label{lemma_asymptotic_value}
  Under the same condition as Lemma \ref{lemma_boresight},
  the asymptotic SNR as the XL-IRS size goes to infinity can be
  expressed as \eqref{eq_snr_special_asymptotic_integral_with relation},
  shown at the top of the next page.
\end{lemma}

\begin{IEEEproof}
  As the IRS size $L_y,L_z \rightarrow \infty$,
  the radii of the inscribed disk and circumscribed disk
  of the IRS region $L_y \times L_z$
  defined in \eqref{eq_inscribed_with_circumscribed} also go to infinity,
  i.e., $R_1,R_2 \rightarrow \infty$.
  It can be proved that the SNR lower- and upper-bounds 
  given by Lemma \ref{lemma_boresight}
  approach to the same limit due to the identical form of the function $G(R,q\prime)$
  with $R \rightarrow \infty$.
  Therefore,
  the asymptotic SNR in \eqref{eq_snr_special_asymptotic_integral_with relation}
  can be obtained according to the Squeeze Theorem
  \cite{thomas1961calculus}.
\end{IEEEproof}

It can be shown that 
the asymptotic SNR is only determined by the directivity parameter $q\prime$
and the distance ratio $\rho$.
Furthermore, 
it is worth mentioning that 
the convergence of SNR expression 
depends on the integral 
in \eqref{eq_snr_special_asymptotic_integral_with relation}.

In particular, 
for the special case of $\rho=1$, we have 
\begin{equation}
  \label{eq_asymptotic_SNR_rho_1}
  \lim_{L_y,L_z \rightarrow \infty} \gamma = 
  \frac{\lambda^4}{64\pi^2 d^4}
  \left(2+\frac1{q\prime}\right)^2 \bar{P}.
\end{equation}

To gain more insights, 
we will also discuss some special cases for $q\prime$ 
and derive closed-form expressions for the asymptotic SNR.
Firstly, the asymptotic SNR under $q\prime = \frac12$
in \eqref{eq_snr_special_asymptotic_integral_with relation} is 
\begin{equation} 
  \label{eq_snr_special_asymptotic_1}
  \begin{aligned}
    \lim_{L_y,L_z \rightarrow \infty} \gamma =
    \begin{cases}
      \frac{\rho}{1-\rho^2} \frac{\lambda^4}{\pi^2 d^4} \bar{P}
      \left[F\left(\frac12 \arctan \frac{\sqrt{1-\rho^2}}{\rho}\bigg|2\right)\right]^2,&0<\rho<1\\
      \\
      \frac{\lambda^4}{4\pi^2 d^4} \bar{P},&\rho=1
    \end{cases}.
  \end{aligned}
\end{equation} 

Besides, the asymptotic SNR under $q\prime=1$ 
in \eqref{eq_snr_special_asymptotic_integral_with relation} is 
  \begin{equation} 
    \label{eq_snr_special_asymptotic_2}
    \lim_{L_y,L_z \rightarrow \infty} \gamma =
    \begin{cases}
      \frac{\rho^2 (\ln \rho)^2}{(1-\rho^2)^2}
      \frac{9 \lambda^4}{4 \pi^2 d^4} \bar{P},&0<\rho<1\\
      \\
      \frac{9 \lambda^4}{64 \pi^2 d^4} \bar{P},&\rho=1
  \end{cases}.
\end{equation}

Note that 
the asymptotic SNR in \eqref{eq_asymptotic_SNR_rho_1}
will go to infinity by letting $q\prime=0$, 
which is expected since modelling each reflecting element semi-isotropically 
would lead to unbounded power when the XL-IRS size goes to infinity. 
By contrast,
the SNR with considering non-isotropic directional gain pattern 
of IRS's reflecting elements,
e.g., $q\prime =\frac12$ or $q\prime=1$,
will lead to a constant value
in \eqref{eq_snr_special_asymptotic_1} and \eqref{eq_snr_special_asymptotic_2}
as the IRS size increases,
which only depends on the distance ratio $\rho$.
This implies that for XL-IRS-aided communication system,
it is necessary to take into account the impact of the directivity of 
each reflecting element
so as to obey the law of power conservation.

\subsection{Conventional UPW Model}

As a comparison,
the conventional UPW model that also takes into account the directional gain pattern 
of IRS's reflecting elements
can be expressed as
\begin{equation} 
  \label{eq_square_power_scaling_law}
  \begin{aligned}
    \gamma_{UPW}
    =\frac{M^2 \beta_0^2 \bar{P}}{r_q^2 r_p^2}
    G_e\left(\theta_q,\phi_q\right) AF\left(\theta_q,\phi_q\right)
    \times G_e\left(\theta_p,\phi_p\right) AF\left(\theta_p,\phi_p\right),
  \end{aligned}
\end{equation}
which follows based on the approximation that 
all reflecting elements share approximately the same AoA/AoD and link distances;
$\beta_0$ is the channel gain
at the reference distance of 1 meter (m), 
$G_e(\theta,\phi)$ accounts for the directional gain pattern,
and $AF(\theta,\phi)$ accounts for the \emph{array factor}\cite{balanis2015antenna}, 
expressed as
\begin{equation}
  \label{eq_array_factor}
  AF(\theta,\phi)= \frac1{M}
  \frac{\sin\left(\frac{M_y}2 kd\Phi\right)}{\sin\left(\frac12 kd\Phi\right)}
  \frac{\sin\left(\frac{M_z}2 kd\Theta\right)}{\sin\left(\frac12 kd\Theta\right)},
\end{equation}
where $k=\frac{2\pi}{\lambda}$ is the wave number; 
$\Phi = \sin\theta \sin\phi$,
$\Theta = \cos\theta$.

Note that the first part of the SNR expression \eqref{eq_square_power_scaling_law} 
is well known as 
the square power scaling law 
for IRS-assisted communication \cite{wu2019intelligent},
where the SNR increases linearly with the square of
the number of IRS reflecting elements, i.e., $M^2$.
It can be observed that 
such a fundamental result \eqref{eq_square_power_scaling_law}
is also related with the AoA/AoD 
via the directional gain pattern $G_e(\theta,\phi)$ 
and the array factor $AF(\theta,\phi)$.
Furthermore,
it is worth mentioning that 
the above result is valid only when both link distances $r_q$ and $r_p$ 
are sufficiently large by comparison with the IRS dimension, 
i.e., $M$ is moderately large.
This makes it possible that 
the far-field propagation model can
apply to both the whole IRS as well as its each individual element. 
However, when $M$ is extremely large, 
the conventional model may no more hold
since the square power scaling law reveals that 
the SNR would correspondingly increase without any limited bound, 
which obviously violates the law of power conservation. 
Conversely, our new result shows that 
under the practical NUSW model, 
the SNR increases with $M$,
but with a diminishing return, 
and eventually converge to a constant 
that depends on the distance ratio $\rho$, 
as well as the directivity parameter $q\prime$ 
of each reflecting element.

\begin{figure*}[!t]
  \normalsize
  \setcounter{MYtempeqncnt}{\value{equation}}
  \setcounter{equation}{\value{MYtempeqncnt}}
    
  \begin{equation}
    \label{eq_snr_ula_integral}
    \begin{aligned}
      \gamma
      \simeq
      \frac{A^2 \bar{P} \Psi_q \Psi_p}{16 \pi^2 d^2 r_q^2 r_p^2}
      \left|
      \int_{-\frac{L_z}2}^{\frac{L_z}2} 
      \frac{\mathrm{d} z}
      {\left[\left(1-\frac2{r_q} z\cos\theta_{q}+\frac{z^2}{r_q^2}\right)
      \left(1-\frac2{r_p} z\cos\theta_p+\frac{z^2}{r_p^2}\right)\right]^{(q\prime+1)/2}} \right|^2.
    \end{aligned}
  \end{equation}
    
  \setcounter{equation}{\value{MYtempeqncnt}+1}
  \hrulefill
  \vspace*{4pt}
\end{figure*}

\subsection{ULA-based XL-IRS}
To gain more insights, 
we further consider the special case of ULA-based XL-IRS, 
i.e., $M_y=1$ and $M_z=M$.
In this case,
by substituting $y=0$ and $\mathrm{d} y=d$ into \eqref{eq_snr_integral},
the SNR expression can be reduced to a simpler form as \eqref{eq_snr_ula_integral},
shown at the top of the page.

\begin{lemma}
  \label{leamma_ula_based_snr}
  For wireless communication assisted by ULA-based XL-IRS, 
  when the link distances satisfy $r_q \ll r_p$ (i.e., $\rho \rightarrow 0$), 
  the resulting SNR in \eqref{eq_snr_ula_integral} under 
  the cosine gain pattern based on projected aperture,
  i.e., $q\prime=\frac12$,
  can be expressed in closed-form as 
  \begin{equation} 
    \label{eq_snr_ula_closed_form}
    \gamma=
    \frac{\lambda^4 \bar{P} \Psi_p \cos\phi_q }{4 \pi^4 d^2r_{p}^2} 
    \left[F\left(\frac{\alpha_1}{2}\bigg| 2\right)+F\left(\frac{\alpha_2}{2}\bigg|2\right)\right]^2,
  \end{equation}
  where $\alpha_1=\arctan\frac{L_z/2+r_q\cos\theta_q}{r_q\sin\theta_q}$
  and $\alpha_2=\arctan\frac{L_z/2-r_q \cos\theta_q}{r_q\sin\theta_q}$.
\end{lemma}

\begin{IEEEproof}
  The proof follows the similar steps as Appendix B of \cite{feng2021wireless}, 
  which is omitted for brevity. 
\end{IEEEproof}

It is worth mentioning that 
the additional condition $r_q\ll r_p$ given by Lemma \ref{leamma_ula_based_snr}
is consistent with 
the commonly used IRS deployment strategy
\cite{wu2021intelligentTutorial,lu2021aerial}.
Specifically,
it has been demonstrated that 
IRS should be deployed either close to the BS 
or to the user for maximizing the received SNR. 
Lemma \ref{leamma_ula_based_snr} shows that 
with the developed near-field model, 
the resulting SNR for the special case of ULA-based XL-IRS
depends on the IRS size $L_z$, the link distance $r_q$ and the AoA $\theta_q$,
and is in general
expressed in terms of the two geometric parameters, $\alpha_1$ and $\alpha_2$,
which are the angles 
formed by the line segments
connecting the BS location
and its projection to the IRS,
as well as the two ends of the IRS,
as illustrated in Fig. \ref{IRS_ULA_model}.
In particular, $\alpha_1+\alpha_2$ 
is termed as the \emph{angular span}
\cite{lu2020how}.
It is observed that for any AoA $\theta_q$,
both $\alpha_1$ and $\alpha_2$ increase with the IRS size $L_z$
and decrease with the link distance $r_q$ between the BS and the XL-IRS center.
Due to the fact that the Elliptic Integral function 
$F(\vartheta|2)$ 
monotonically increases with $\vartheta$,
the resulting SNR $\gamma$ in \eqref{eq_snr_ula_closed_form}
increases with $L_z$
but decreases with $r_q$, 
as expected.
Furthermore, 
Lemma \ref{leamma_ula_based_snr} shows that under the practical NUSW model, 
the SNR increases with $L_z$ with a diminishing return.
Note that this result 
differs from the conventional square power scaling law obtained 
based on the UPW model, 
where the SNR increases unboundedly with the IRS size
\cite{wu2019intelligent,lu2021aerial}.
In particular, 
it can be shown that as $L_z \rightarrow \infty$, 
we reach the extreme case, i.e., $\alpha_1=\alpha_2=\frac{\pi}2$,
leading to the following lemma.

\begin{lemma}
  Under the same condition as Lemma \ref{leamma_ula_based_snr}, 
  the asymptotic SNR for ULA-based XL-IRS is 
  \begin{equation}
    \label{eq_ula_based_snr}
    \begin{aligned} 
      \lim_{L_z \rightarrow \infty} \gamma
      =
      \frac{\lambda^4 \bar{P} \Psi_p \cos\phi_q }{\pi^4 d^2 r_p^2} 
      \left[F\left(\frac{\pi}{4}\bigg| 2\right) \right]^2 
      = 1.7188 \times 
      \frac{\lambda^4 \bar{P} \Psi_p}{\pi^4 d^2r_p^2} \cos\phi_q.
    \end{aligned}
  \end{equation}
\end{lemma}

\section{Extension to Multi-Antenna BS}
\begin{figure}[ht]
  \centering
  \includegraphics[width=4.0in]{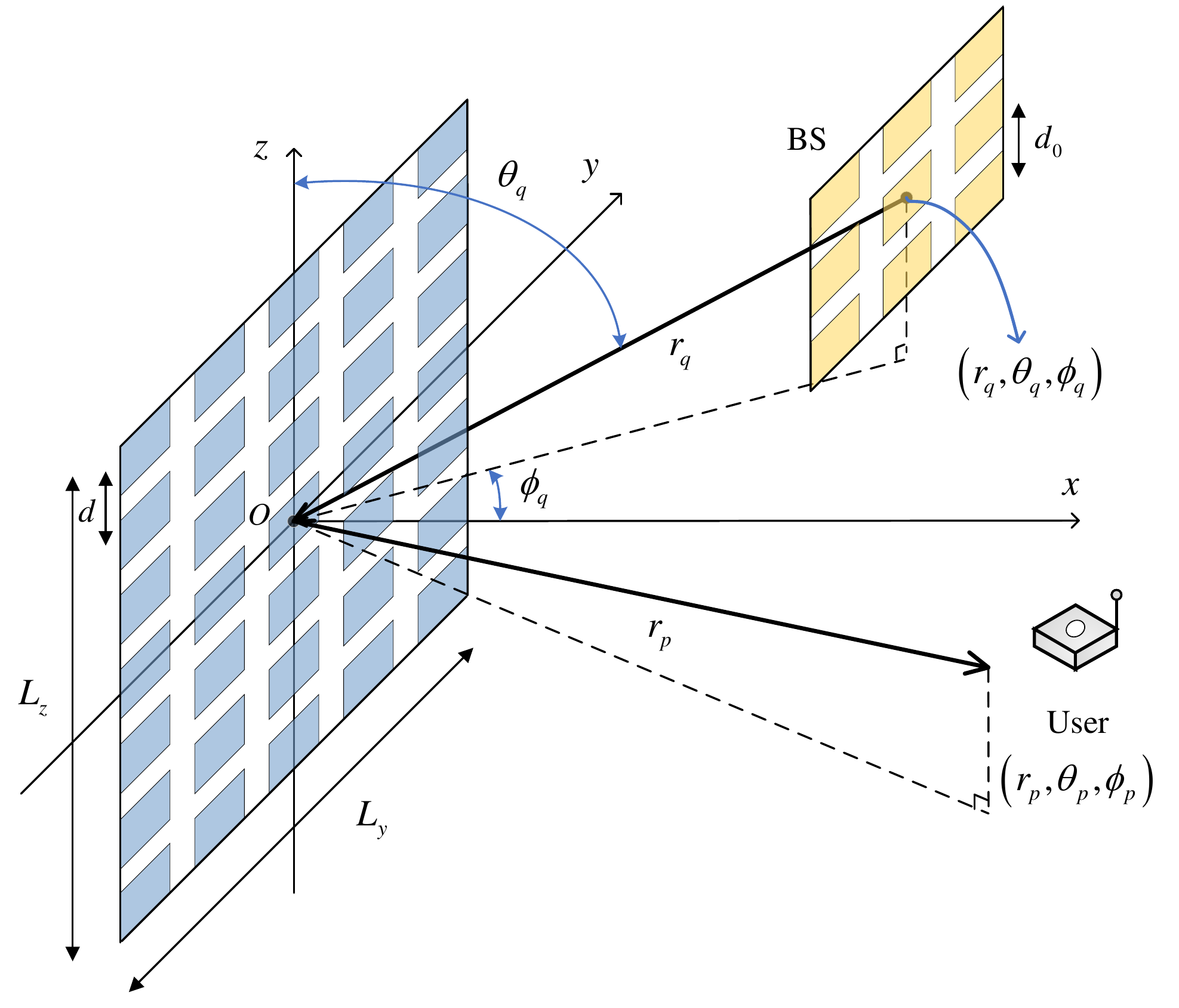}
  \caption{Wireless communication assisted by XL-IRS with multi-antenna BS.}
  \label{XL-IRS-assisted MIMO communication}
\end{figure}

Our previous analysis assumes that 
the BS has a single antenna 
so as to reveal the most important insights. 
In this section, 
we consider the MISO setup where the BS is equipped with multiple antennas 
in the form of UPA architecture. 
For simplicity, 
we assume that the UPA at the BS is parallel
to the IRS,
as illustrated in Fig. \ref{XL-IRS-assisted MIMO communication}. 
The number of antennas at the BS
is denoted as $N=N_{y} N_{z}$, 
where $N_{y}$ and $N_{z}$
denote the number of antennas 
along the $y$- and $z$-axis, respectively,
and the antenna element separation is $d_0$.
Thus, 
the central location of the $\left(n_y,n_z\right)$-th antenna is  
$\mathbf{u}_{n_y,n_z}=\left[r_q \Psi_q,r_q \Phi_q+n_y d_0,r_q \Theta_q+n_z d_0\right]^T$,
where $n_y=0,\pm 1,\cdots,\pm \left(N_y-1\right)/2$ 
and $n_z=0,\pm 1,\cdots,\pm \left(N_z-1\right)/2$.
The distance between the $\left(n_y,n_z\right)$-th antenna at the BS 
and the $\left(m_y,m_z\right)$-th reflecting element is 
\begin{equation}
  \label{eq_distance_BS_IRS_MISO}
    r_{q,m_y,m_z,n_y,n_z}
    =\left\|\mathbf{u}_{m_y,m_z}-\mathbf{w}_{m_y,m_z}\right\|,
\end{equation}
which can be further expressed as \eqref{eq_distance_UPA_long} 
at the top of the next page
with $\varepsilon_{q_0} \triangleq \frac{d_0}{r_q} \ll 1$
\cite{lu2021communicating,feng2021wireless}.

\begin{figure*}[!t]
  \normalsize
  \setcounter{MYtempeqncnt}{\value{equation}}
  \setcounter{equation}{\value{MYtempeqncnt}}

  \begin{equation}
    \label{eq_distance_UPA_long}
  \begin{aligned}
  &r_{q,m_y,m_z,n_y,n_z}\\
  &\quad=r_q \sqrt{1+2\Phi_q(n_y \varepsilon_{q_0}-m_y \varepsilon_q)
  +2\Theta_q(n_z \varepsilon_{q_0}-m_z \varepsilon_q)
  +(n_y \varepsilon_{q_0}-m_y \varepsilon_q)^2
  +(n_z \varepsilon_{q_0}-m_z \varepsilon_q)^2}.
  \end{aligned}
  \end{equation}

  \setcounter{equation}{\value{MYtempeqncnt}+1}
  \hrulefill
  \vspace*{4pt}
\end{figure*}
Thus, 
the channel matrix between the BS and XL-IRS is denoted as 
$\mathbf{H}\in \mathbb{C}^{M\times N}$,
whose entries are given by 
\begin{equation}
  \label{eq_channel_vector_BS_MISO}
  \begin{aligned}
    h_{m_y,m_z,n_y,n_z}
    =\sqrt{a_{m_y,m_z,n_y,n_z}\left(r_q,\theta_q,\phi_q\right)}
    \times e^{-j\frac{2\pi}{\lambda}r_{q,m_y,m_z,n_y,n_z}},
    \forall m_y,m_z,n_y,n_z,
  \end{aligned}
\end{equation}
where the channel power gain $a_{m_y,m_z,n_y,n_Z}(r_q,\theta_q,\phi_q)$
between the $(n_y,n_z)$-th antenna at the BS 
and the $(m_y,m_z)$-th reflecting element at the IRS is 
given by \eqref{eq_power_gain_BS_MISO} at the top of the page.
\begin{figure*}[!t]
  \normalsize
  \setcounter{MYtempeqncnt}{\value{equation}}
  \setcounter{equation}{\value{MYtempeqncnt}}

  \begin{equation}
    \label{eq_power_gain_BS_MISO}
  \begin{aligned}
  &a_{m_y,m_z,n_y,n_Z}
  (r_q,\theta_q,\phi_q)\\
  &=\left(\frac{\lambda}{4\pi r_q}\right)^2 
  \frac{\gamma\prime \Psi_q^{2q\prime}}
  {\left[1+2(n_y \varepsilon_{q_0}-m_y \varepsilon_q) \Phi_q
  +2(n_z \varepsilon_{q_0}-m_z \varepsilon_q) \Theta_q
  +(n_y \varepsilon_{q_0}-m_y \varepsilon_q)^2
  +(n_z \varepsilon_{q_0}-m_z \varepsilon_q)^2
  \right]^{q\prime+1}}.
  \end{aligned}
  \end{equation}

  \setcounter{equation}{\value{MYtempeqncnt}+1}
  \hrulefill
  \vspace*{4pt}
  \end{figure*}

Therefore,
the SNR at the user can be expressed as 
\begin{equation}
  \label{eq_received_snr_MISO}
  \gamma=\bar{P}\left|\mathbf{g}^T \mathbf{\Theta} \mathbf{H}\mathbf{v}\right|^2,
\end{equation}
where $\mathbf{v}\in\mathbb{C}^{N \times 1}$ 
is a beamforming vector, 
with $\|\mathbf{v}\|=1$.

Similarly,
it is assumed that
the IRS is deployed at the vicinity of the user,
so that 
the user is located in the
near-field region of the IRS while the BS is in the far-
field region.
By letting $\varepsilon_q \rightarrow 0$
in \eqref{eq_power_gain_BS_MISO},
the channel matrix vector $\mathbf{H} \in \mathbb{C}^{M \times N}$
can be written as a rank-one matrix as 
\begin{equation}
  \label{eq_far_vector_matrix}
  \mathbf{H}=
  \sqrt{\frac{\lambda^2 \gamma\prime \Psi_q^{2q'}}{16\pi^2 r_q^2}} 
  e^{-j\frac{2\pi}{\lambda}r_q}
  \mathbf{a}_{R}\left(\theta_q,\phi_q\right)
  \mathbf{a}_{T}^H\left(\theta_q,\phi_q\right),
\end{equation}
where $\mathbf{a}_{R}\left(\theta_q,\phi_q\right)\in \mathbb{C}^{M \times 1}$ 
is the receive array response vector at the XL-IRS,
given by
\begin{equation}
  \label{eq_receiver_array}
  \mathbf{a}_{R}(\theta_q,\phi_q)
  =\left[e^{j\frac{2\pi}{\lambda}m_y d\Phi_q}\right]_{1\times M_y}^T
  \otimes \left[e^{j\frac{2\pi}{\lambda}m_z d\Theta_q}\right]_{1\times M_z}^T,
\end{equation}
where the symbol $\otimes$ denotes the Kronecker product.
Similarly,
denote the transmit array response vector at the BS as
$\mathbf{a}_{T}\left(\theta_q,\phi_q\right) \in \mathbb{C}^{N \times 1}$,
given by
\begin{equation}
  \label{eq_transmit_array}
  \mathbf{a}_{T}(\theta_q,\phi_q)
  =\left[e^{j\frac{2\pi}{\lambda}n_y d_0\Phi_q}\right]_{1\times N_y}^T
  \otimes \left[e^{j\frac{2\pi}{\lambda}n_z d_0\Theta_q}\right]_{1\times N_z}^T.
\end{equation}

Based on \cite{lu2021aerial},
the optimal transmit beamforming vector 
at the BS for SNR maximization is
\begin{equation}
  \label{eq_optimal_beamforming}
  \mathbf{v}^*=\frac{\mathbf{a}_{T}\left(\theta_q,\phi_q\right)}{\sqrt{N}}.
\end{equation}

Furthermore, with the optimal phase shifting 
by the XL-IRS in \eqref{eq_received_snr},
the maximum SNR at the user reduces to
\begin{equation}
  \label{eq_far_max_snr}
  \gamma=N\bar{P}
  \left|\sum_{m_y}\sum_{m_z} 
  \sqrt{\frac{\lambda^2 \gamma\prime \Psi_q^{2q'}}{16\pi^2 r_q^2}b_{m_y,m_z}\left(r_p,\theta_p,\phi_p\right)}\right|^2.
\end{equation}
By substituting \eqref{eq_power_gain_RX} into \eqref{eq_far_max_snr},
we obatin the summation form of the SNR at the user 
as \eqref{eq_far_max_snr_sum}, shown at the top of the page.

\begin{figure*}[!t]
  \normalsize
  \setcounter{MYtempeqncnt}{\value{equation}}
  \setcounter{equation}{\value{MYtempeqncnt}}
  
  \begin{equation}
    \label{eq_far_max_snr_sum}
    \begin{aligned}
      \gamma=N \bar{P} 
      \left(\frac{\lambda}{4\pi}\right)^4
      \frac{\gamma\prime^2 \Psi_q^{2q'} \Psi_p^{2q'}}{r_q^2 r_p^2}
      \bigg|\sum_{m_y,m_z} 
      \frac1{\left[1-2m_y\varepsilon_p\Phi_p-2m_z\varepsilon_p\Theta_p+(m_y^2+m_z^2)\varepsilon_p^2\right]^{(q'+1)/2}}
      \bigg|^2.
    \end{aligned}
  \end{equation}
  
  \setcounter{equation}{\value{MYtempeqncnt}+1}
  \hrulefill
  \vspace*{4pt}
\end{figure*}

\begin{figure*}[!t]
  \normalsize
  \setcounter{MYtempeqncnt}{\value{equation}}
  \setcounter{equation}{\value{MYtempeqncnt}}
    
  \begin{equation}
    \label{eq_far_max_snr_integral}
    \begin{aligned}
      \gamma=N \bar{P} 
      \left(\frac{\lambda}{4\pi}\right)^4
      \frac{\gamma\prime^2 \Psi_q^{2q'} \Psi_p^{2q'}}{d^4 r_q^2 r_p^2}
      \bigg|\int_{-\frac{L_z}2}^{\frac{L_z}2}
      \int_{-\frac{L_y}2}^{\frac{L_y}2}
      \frac{\mathrm{d} y \mathrm{d} z}{\left[1-\frac2{r_p} y \Phi_p-
      \frac2{r_p} z \Theta_p
      +\frac2{r_p^2}(y^2+z^2)\right]^{(q'+1)/2}}
      \bigg|^2.
    \end{aligned}
  \end{equation}
    
  \setcounter{equation}{\value{MYtempeqncnt}+1}
  \hrulefill
  \vspace*{4pt}
\end{figure*}

Similarly, 
by approximating the summation with a double integral
by using the fact $\varepsilon_p \ll 1$, 
the SNR can be written in an integral form \eqref{eq_far_max_snr_integral}
shown at the top of the page.
Following the similar derivation in Theorem \ref{theorem_bounds}, 
the SNR is bounded by 
\begin{equation}
  \label{eq_far_miso_bounds}
  U(R_1,q\prime) \leq \gamma \leq U(R_2,q\prime),
\end{equation}
where the function $U(R,q\prime)$ is defined as \eqref{eq_far_max_snr_integral_bounds}
shown at the top of the page,
and the radii $R_1$ and $R_2$ are also given by \eqref{eq_inscribed_with_circumscribed}.

\begin{figure*}[!t]
  \normalsize
  \setcounter{MYtempeqncnt}{\value{equation}}
  \setcounter{equation}{\value{MYtempeqncnt}}
    
  \begin{equation}
    \label{eq_far_max_snr_integral_bounds}
    \begin{aligned}
      U(R,q\prime) \triangleq N \bar{P} 
      \left(\frac{\lambda}{4\pi}\right)^4
      \frac{\gamma\prime^2 \Psi_q^{2q'} \Psi_p^{2q'}}{d^4 r_q^2 r_p^2}
      \bigg|\int_0^{2\pi} \mathrm{d} \zeta
      \int_0^R
      \frac{r\mathrm{d} r}{\left[1-\frac{2r}{r_p} \Phi_p\cos\zeta 
      -\frac{2r}{r_p} \Theta_p\sin\zeta
      +\frac{r^2}{r_p^2}\right]^{(q'+1)/2}}
      \bigg|^2.
    \end{aligned}
  \end{equation}
    
  \setcounter{equation}{\value{MYtempeqncnt}+1}
  \hrulefill
  \vspace*{4pt}
\end{figure*}
       
\begin{lemma}
  \label{bounds_MISO_case}
  If the user is located along the boresight 
  of the XL-IRS, i.e., near the $x$-axis with
  $\Phi_p \ll \frac{r_p}{L_y}$, $\Theta_p \ll \frac{r_p}{L_z}$,
  and $q'=\frac12$,
  we have
  \begin{equation}
    \label{eq_far_miso_bounds_function}
    U\left(R,\frac12\right) \triangleq N\bar{P}
    \frac{\lambda^4 \gamma\prime^2 \Psi_q r_p^2}{16 \pi^2 d^4 r_q^2}
    \bigg(\sqrt[4]{\frac{R^2}{r_p^2}+1}-1\bigg)^2.
  \end{equation}
\end{lemma}

\begin{IEEEproof}
  Please refer to Appendix C.
\end{IEEEproof}

\section{Numerical Results}

In this section,
numerical results are provided to 
validate our near-field model and theoretical analysis 
for XL-IRS-aided communications,
and also compare our developed model with the conventional UPW model.
Unless otherwise stated,
the signal wavelength is set as $\lambda=0.125\,$m,
and the separation of adjacent IRS elements 
is $d=\frac{\lambda}3$.

\subsection{SNR Bounds and Asymptotic Analysis}


\begin{figure*}[!t]
  \centering
  \subfloat[$q\prime=0$]{\includegraphics[width=3.0in]{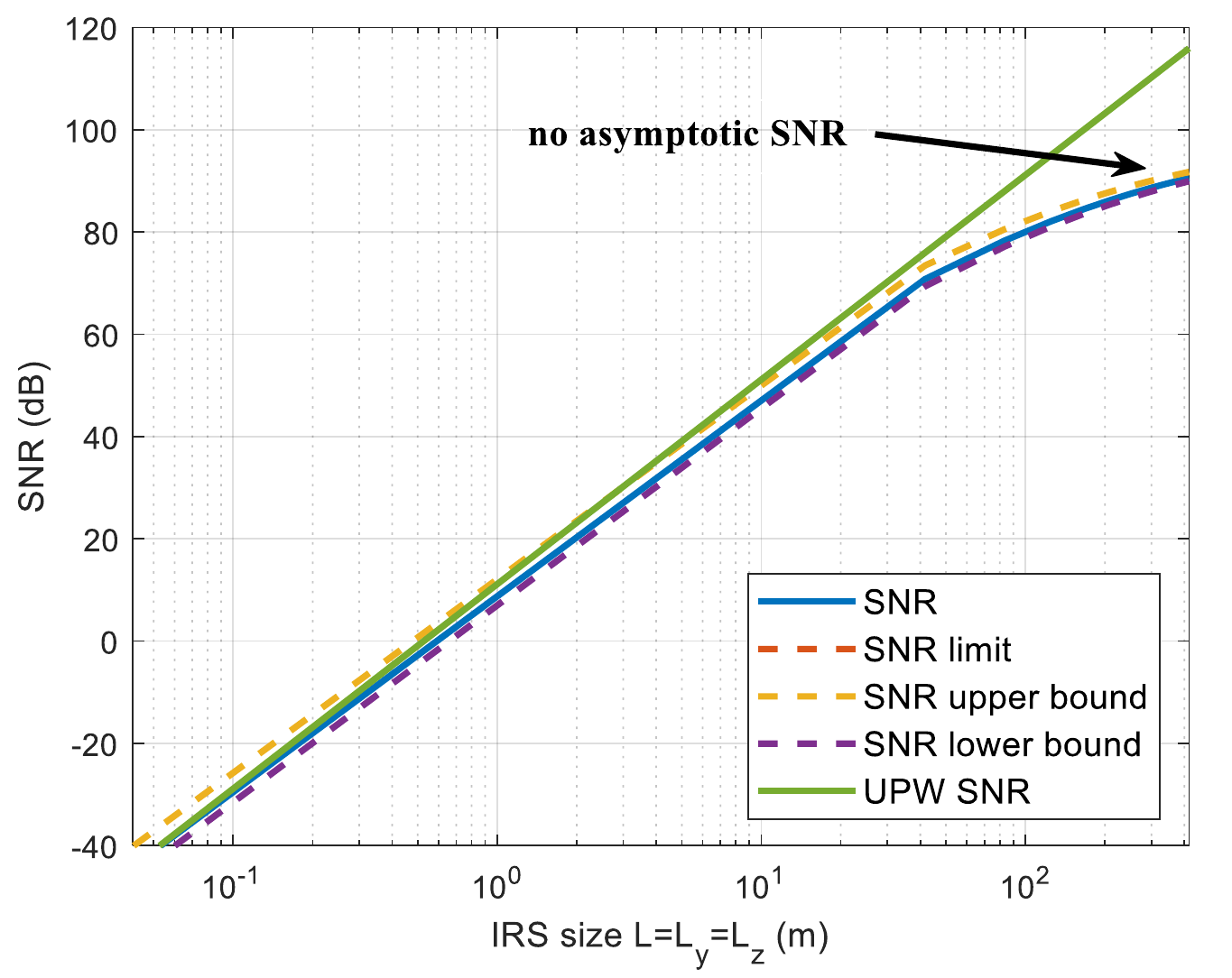}
  \label{snr_numerical_results_IRS_size_q_0}}
  \hfil
  \subfloat[$q\prime=\frac12$]{\includegraphics[width=3.0in]{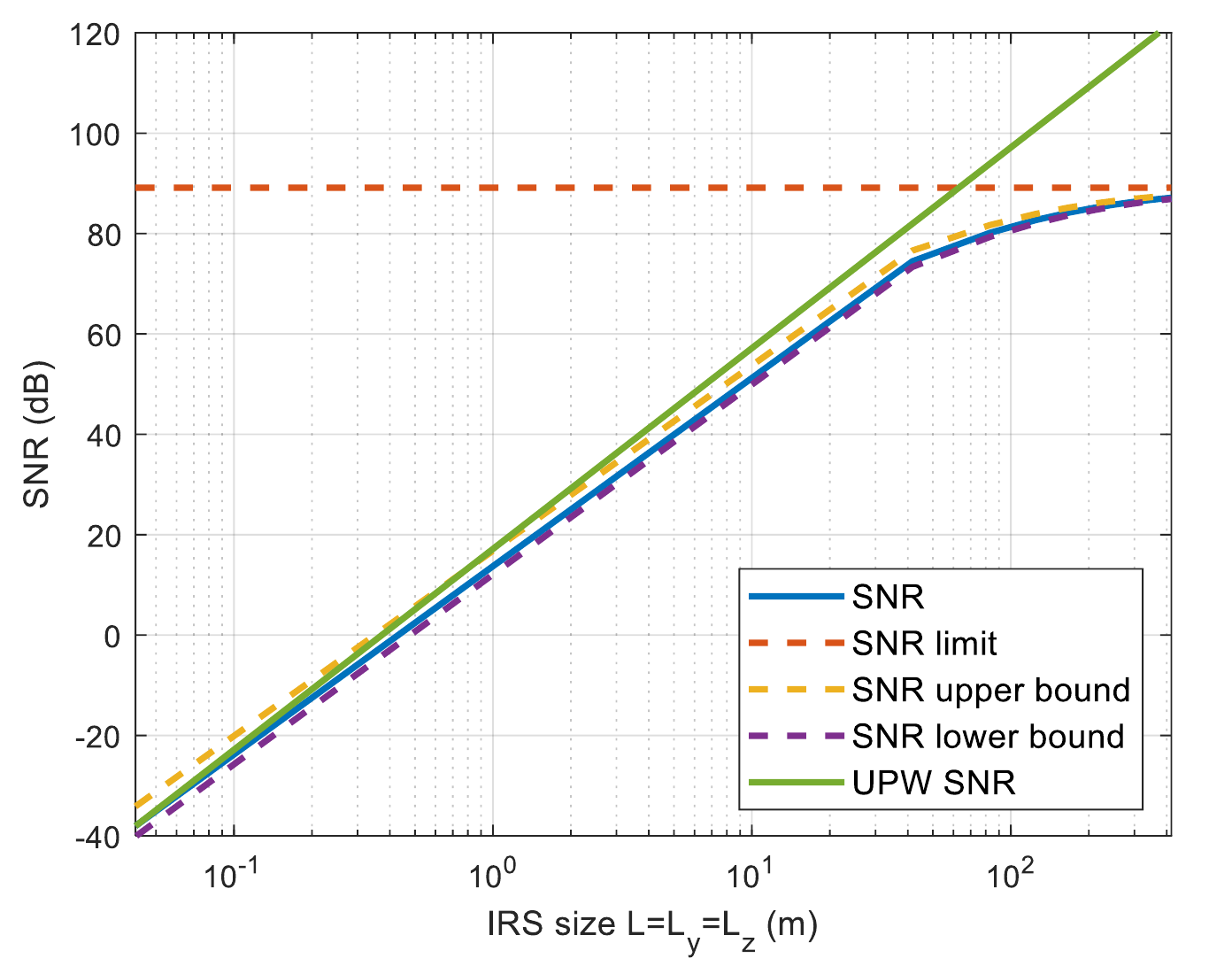}
  \label{snr_numerical_results_IRS_size_q_1/2}}
  \hfil
  \subfloat[$q\prime=1$]{\includegraphics[width=3.0in]{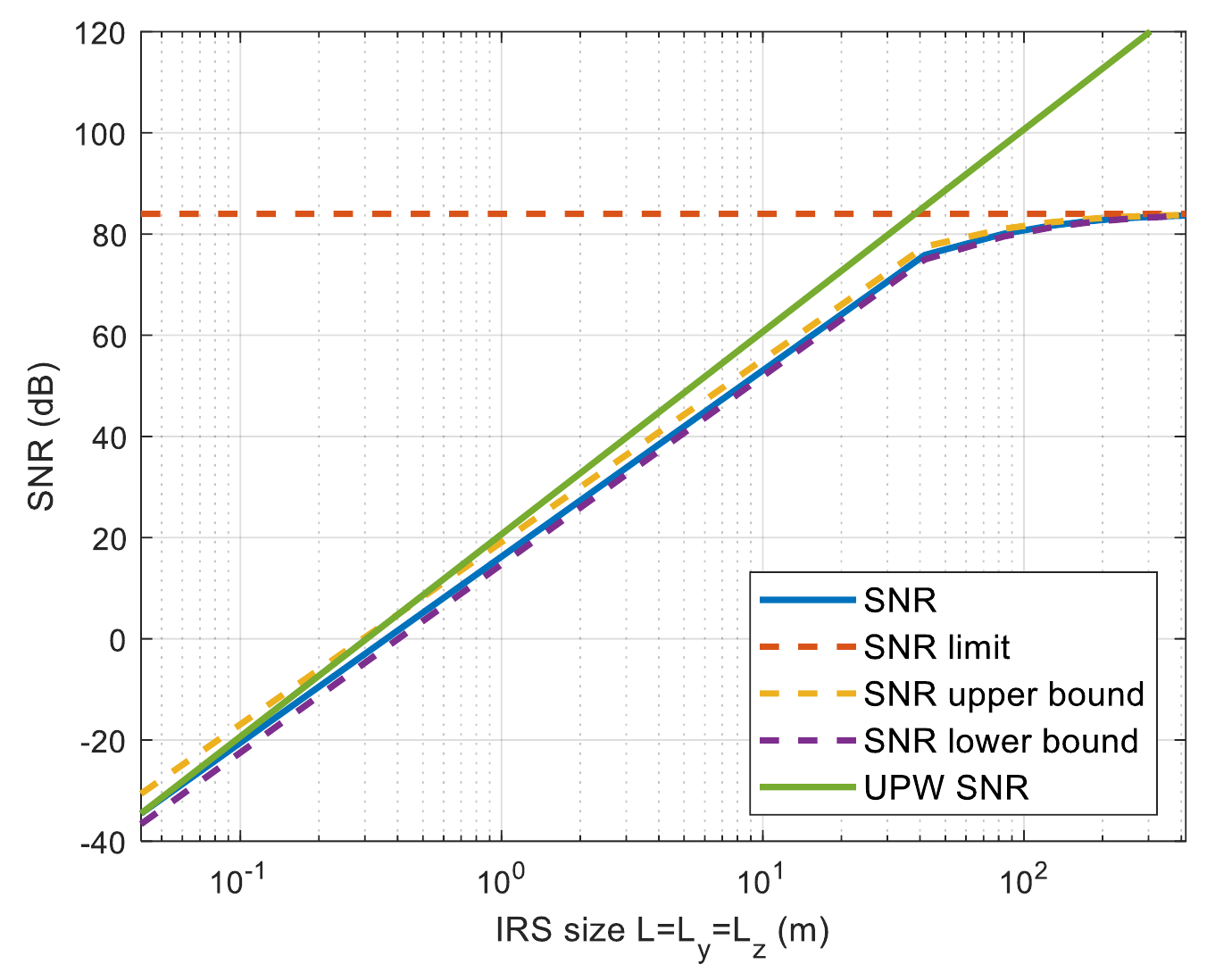}
  \label{snr_numerical_results_IRS_size_q_1}}
  \caption{SNR versus IRS size.}
  \label{snr_numerical_results_IRS_size}
\end{figure*}

Firstly, we present
the numerical results of the SNR bounds and its asymptotic performance.
Fig. \ref{snr_numerical_results_IRS_size} plots the SNR 
versus the IRS size for a square UPA-based XL-IRS,
i.e., $L=L_y=L_z$. 
Three directional gain patterns of IRS's reflecting element
based on
\eqref{eq_element_pattern_istropic}-\eqref{eq_element_pattern_square}
for ease of exposition are considered.
Additionally,
the derived results in the previous sections are compared 
in Fig. \ref{snr_numerical_results_IRS_size},
based on the summation in \eqref{eq_snr_summation},
lower- and upper-bounds
in \eqref{eq_G_integral}-\eqref{eq_snr_special_bounds_2},
the asymptotic value 
in \eqref{eq_snr_special_asymptotic_integral_with relation}, \eqref{eq_snr_special_asymptotic_1}
and \eqref{eq_snr_special_asymptotic_2},
and the conventional UPW model 
in \eqref{eq_square_power_scaling_law}.
The transmit SNR is $\bar{P}=90\,\mathrm{dB}$,
and the BS and user are assumed to be located at 
$\mathbf{q}=[10,0,0]^T\,\mathrm{m}$
and $\mathbf{p}=[100,0,0]^T\,\mathrm{m}$, respectively.
It is firstly observed that for UPA-based XL-IRS,
the derived bounds in Lemma \ref{lemma_boresight}
are sufficiently tight for the SNR estimation.
In particular, 
Figs. \ref{snr_numerical_results_IRS_size}\subref{snr_numerical_results_IRS_size_q_1/2}
and \ref{snr_numerical_results_IRS_size}\subref{snr_numerical_results_IRS_size_q_1}
validate our closed-form bounds
in \eqref{eq_snr_special_bounds_1} and \eqref{eq_snr_special_bounds_2}.
Furthermore,
as the IRS size $L$ becomes large,
the SNR in 
Fig. \ref{snr_numerical_results_IRS_size}\subref{snr_numerical_results_IRS_size_q_0}
goes to infinity eventually,
while approaching to a constant in 
Fig. \ref{snr_numerical_results_IRS_size}\subref{snr_numerical_results_IRS_size_q_1/2}
and \ref{snr_numerical_results_IRS_size}\subref{snr_numerical_results_IRS_size_q_1}.
This validates our theoretical results in Lemma \ref{lemma_asymptotic_value}.
Besides,
it is observed that 
the conventional UPW model 
in \eqref{eq_square_power_scaling_law}
is approximately consistent with our developed near-field model
when the IRS size is not very large.
However, 
as the IRS size exceeds a certain threshold, 
such two models exhibit drastically different scaling laws,
i.e., 
converging to a constant value 
versus increasing unboundedly.
In summary, 
the conventional UPW model is valid for most practical cases
in terms of the received power,
but the accurate near-field modelling
needs to  be considered when the IRS size becomes significantly large,
especially for the asymptotic analysis,
which is consistent with the conclusions in \cite{bjornson2020power}.
\begin{figure*}[!t]
  \centering
  \subfloat[$q\prime=0$]{\includegraphics[width=3.0in]{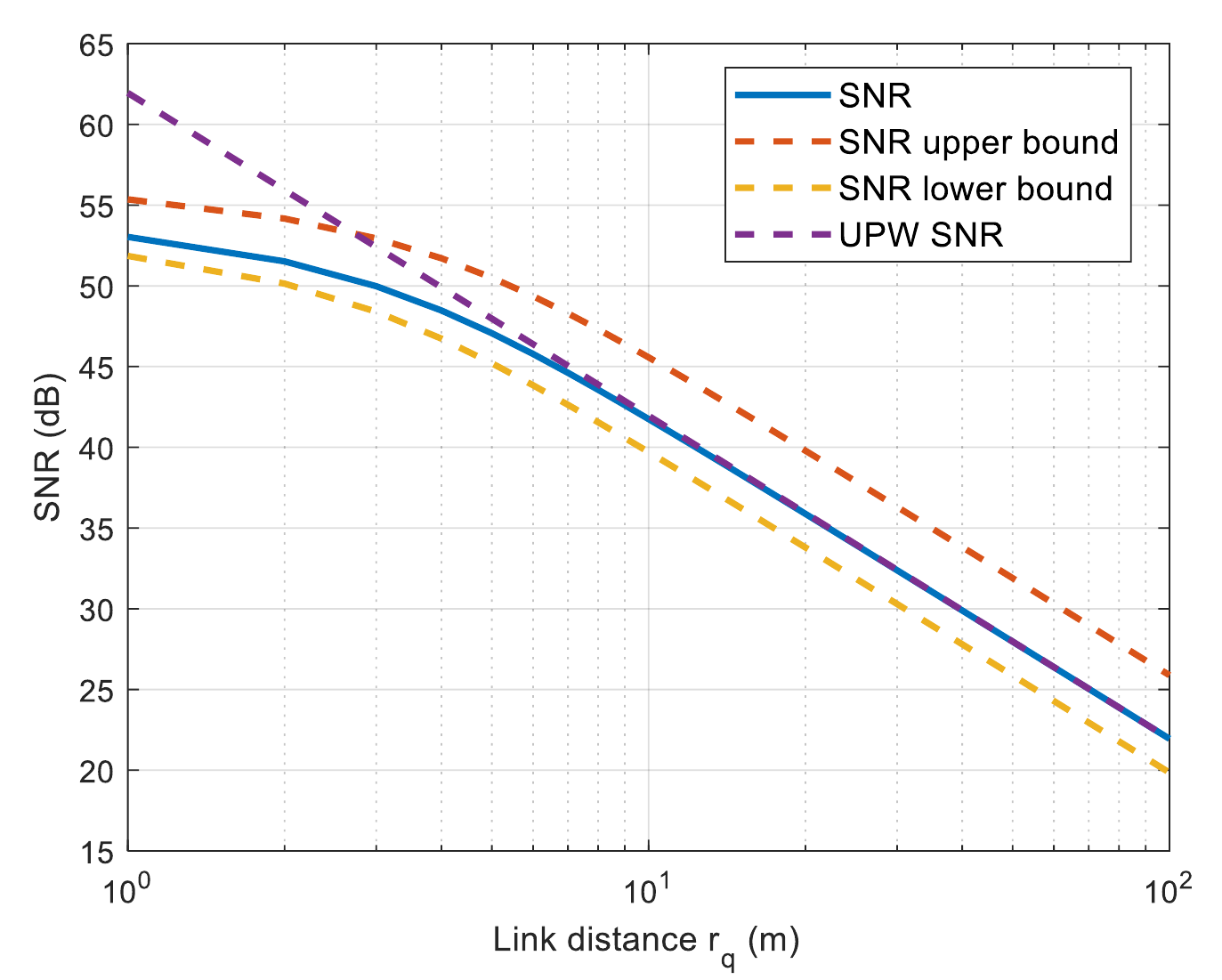}
  \label{snr_numerical_results_link_distance_q_0}}
  \hfil
  \subfloat[$q\prime=\frac12$]{\includegraphics[width=3.0in]{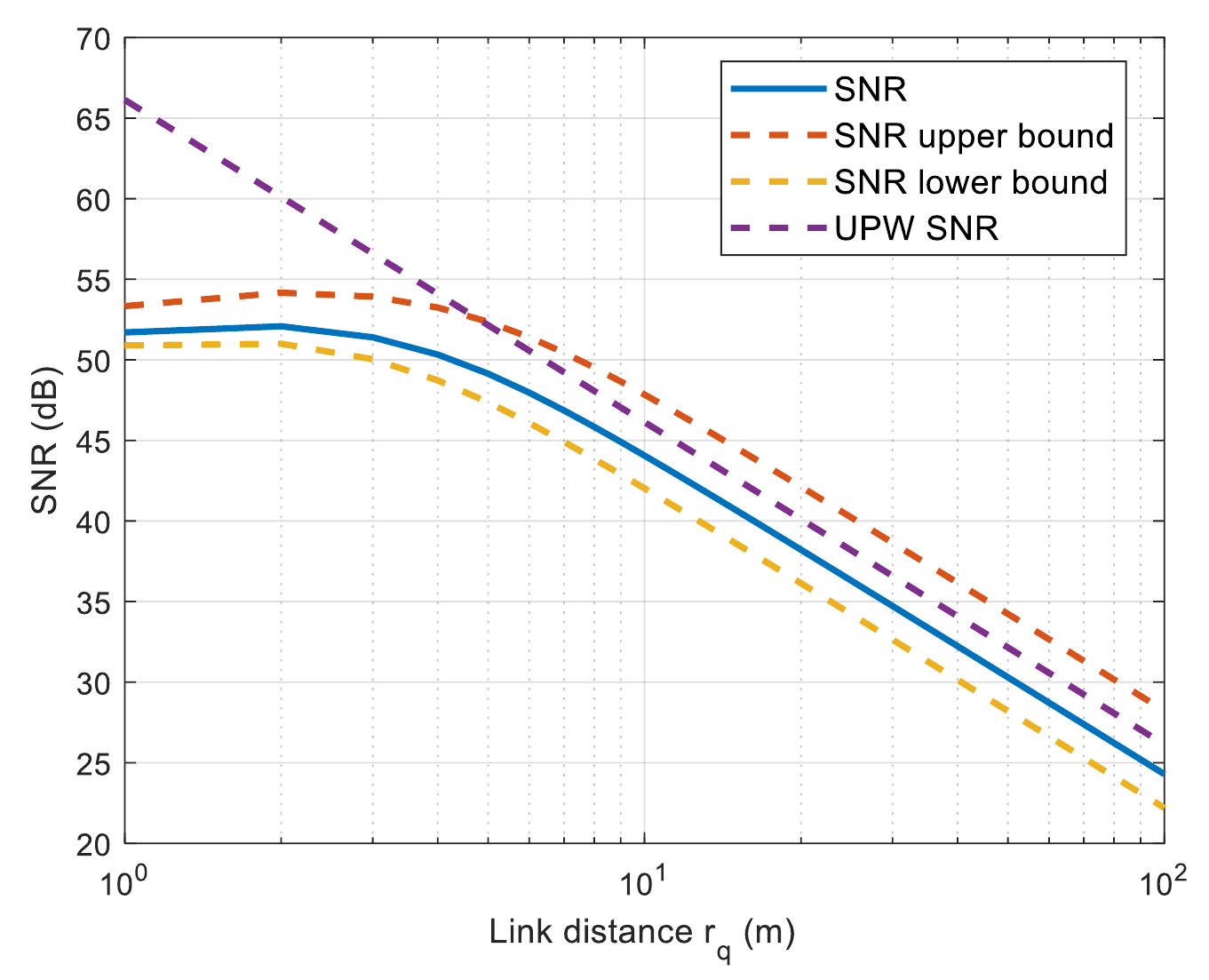}
  \label{snr_numerical_results_link_distance_q_1/2}}
  \hfil
  \subfloat[$q\prime=1$]{\includegraphics[width=3.0in]{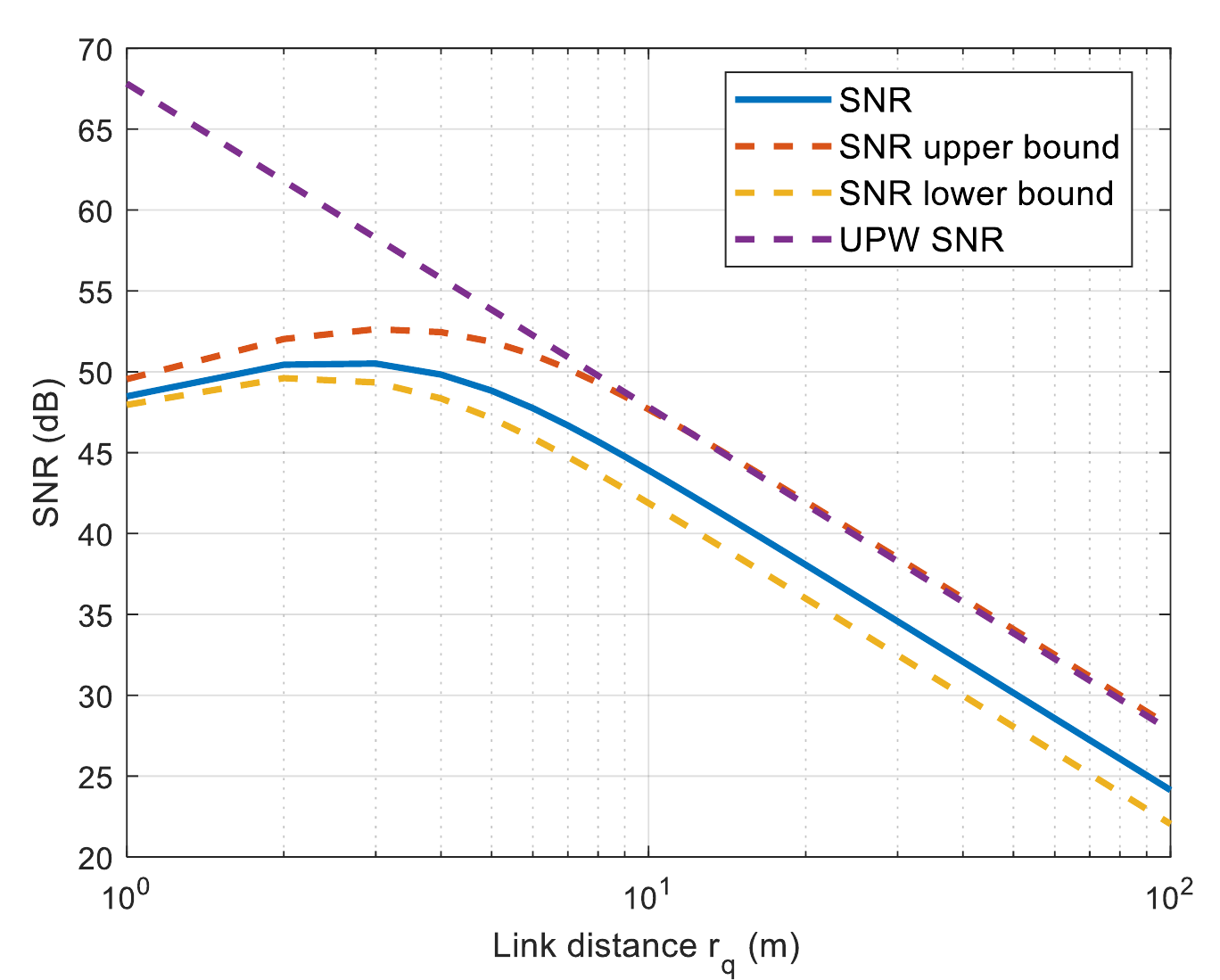}
  \label{snr_numerical_results_link_distance_q_1}}
  \caption{SNR versus link distance $r_q$.}
  \label{snr_numerical_results_link_distance}
\end{figure*}

Fig. \ref{snr_numerical_results_link_distance} plots the SNR 
versus the link distance $r_q$ between the BS and the IRS center,
where the square IRS size is set as $L_y=L_z \simeq 8\,\mathrm{m}$.
We compare the resulting SNR with the following expressions,
i.e.,
the summation form in \eqref{eq_snr_summation},
generic lower- and upper-bounds in \eqref{eq_siso_general_bounds},
and the square power scaling law under the conventional UPW model 
in \eqref{eq_square_power_scaling_law}.
The BS direction is set as
$(\theta_q,\phi_q)=\left(\frac{\pi}3,\frac{\pi}6\right)$,
and the user is located at 
$(r_p,\theta_p,\phi_p)=\left(200\,\mathrm{m},\frac{3\pi}4,-\frac{\pi}5\right)$.
It is observed that 
the SNR bounds given in Theorem \ref{theorem_bounds}
are rather accurate,
and the conventional UPW model in general
over-estimates the SNR values 
when taking into account the directional gain patterns
of IRS's reflecting elements,
as illustrated in 
Fig. \ref{snr_numerical_results_link_distance}\subref{snr_numerical_results_link_distance_q_1/2}
and \ref{snr_numerical_results_link_distance}\subref{snr_numerical_results_link_distance_q_1}.
In particular,
Fig. \ref{snr_numerical_results_link_distance}\subref{snr_numerical_results_link_distance_q_1}
reveals that 
the SNR does not necessarily monotonically decrease with the distance
from the BS to the XL-IRS,
because it also depends on the gain pattern of each element.
Although the BS is deployed close to the IRS,
the smaller AoA from the BS to each reflecting element 
will result in the moderately worse performance 
when its directional gain pattern
is strong in the boresight direction.
Therefore,
as the link distance $r_q$ grows,
there might be first a slight rise for the SNR 
due to the increase of the AoA from the BS to each IRS element.

\begin{figure}[ht]
  \centering
  \includegraphics[width=4.2in]{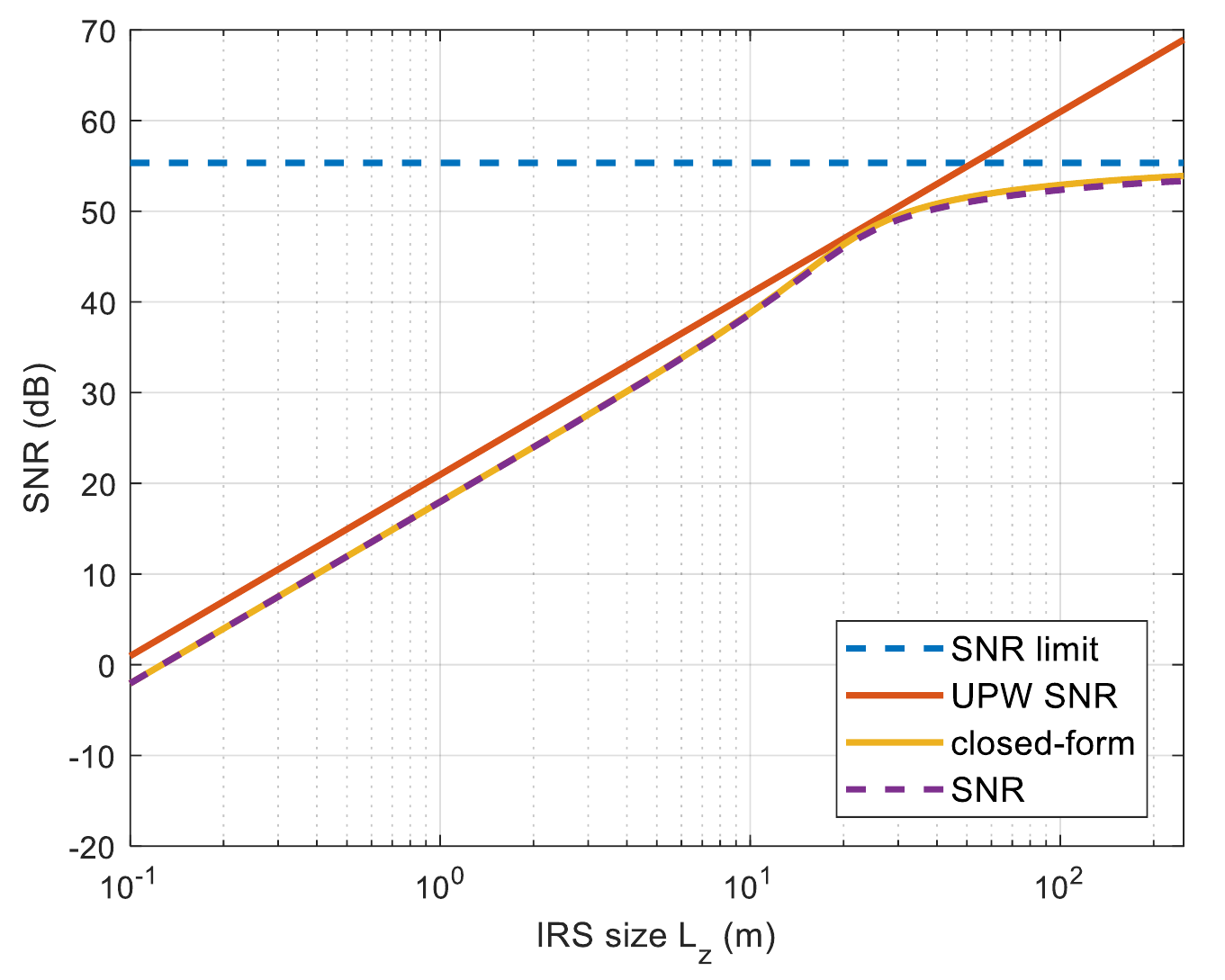}
  \caption{SNR versus IRS size for ULA-based XL-IRS.}
  \label{IRS_ULA_snr_versus_M}
\end{figure}

\subsection{ULA-based XL-IRS}
Next,
for the special case of ULA-based XL-IRS,
Fig. \ref{IRS_ULA_snr_versus_M} plots the SNR
versus the IRS size $L_z$
based on the summation in \eqref{eq_snr_summation},
derived closed-form expression in \eqref{eq_snr_ula_closed_form},
the asymptotic limit in \eqref{eq_ula_based_snr},
and the conventional UPW model in \eqref{eq_square_power_scaling_law}.
The transmit SNR is $\bar{P}=120\,\mathrm{dB}$,
and the BS and user are located at
$(r_q,\theta_q,\phi_q)=\left(10\,\mathrm{m},\frac{\pi}3,\frac{\pi}6\right)$
and
$(r_p,\theta_p,\phi_p)=\left(100\,\mathrm{m},\frac{3\pi}4,-\frac{\pi}5\right)$,
respectively.
It is firstly observed that both
the closed-form expression \eqref{eq_snr_ula_closed_form} and 
the asymptotic limit \eqref{eq_ula_based_snr} 
match quite well with the actual values. 
Furthermore,
similar to Fig. \ref{snr_numerical_results_IRS_size},
the UPW expression can accurately predict 
the SNR values for the moderate-scale IRS.
Likewise,
as $L_z$ goes extremely large,
although the SNR under the conventional UPW model increases unboundedly, 
that under our developed near-field model 
approaches to a constant value specified in \eqref{eq_ula_based_snr}.  
This again demonstrates the importance 
of proper near-field modelling 
for communications assisted by XL-IRS.

\subsection{Multi-Antenna BS}
\begin{figure}[ht]
  \centering
  \includegraphics[width=4.2in]{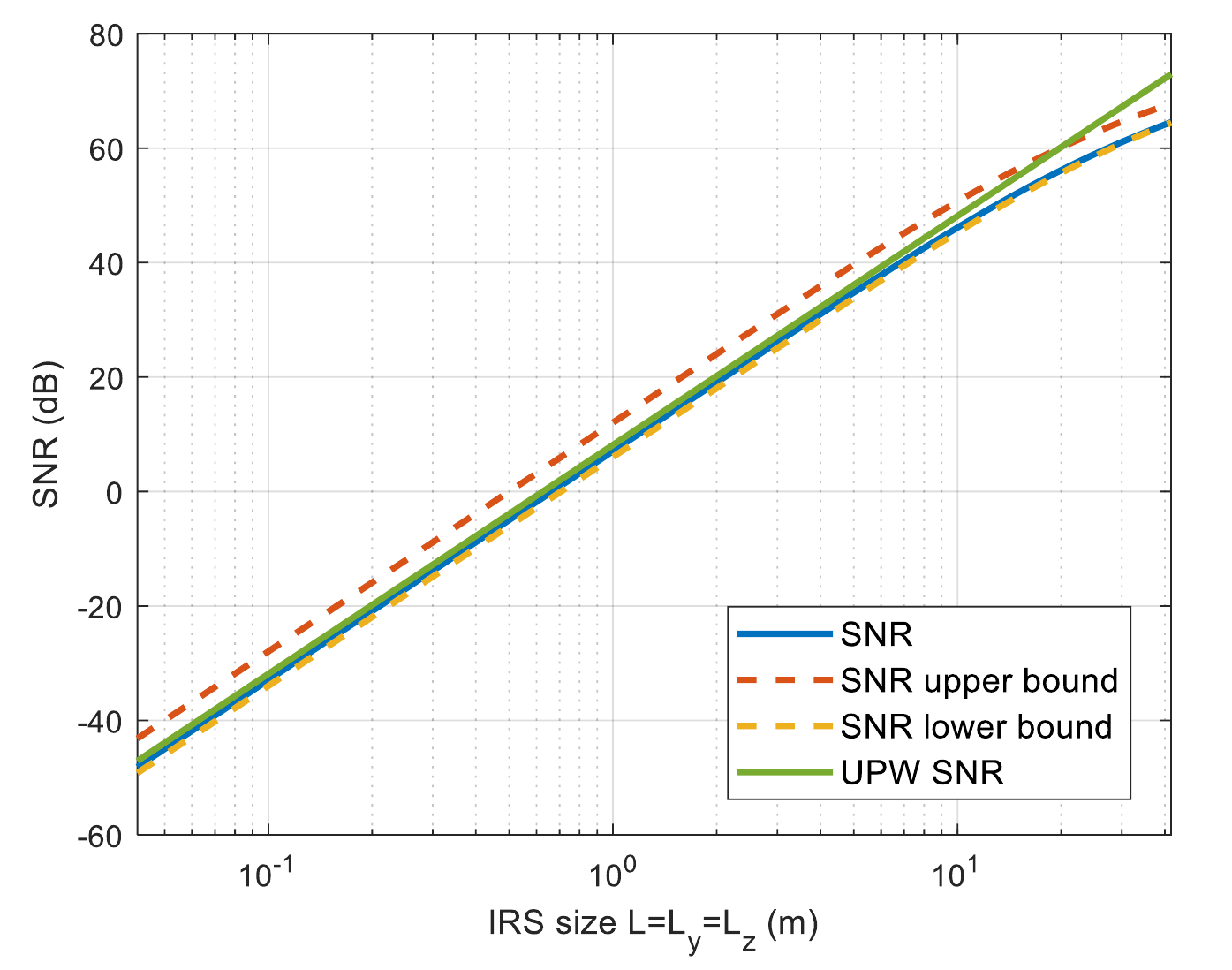}
  \caption{SNR versus IRS size with Multi-Antenna BS.}
  \label{snr_numerical_results_IRS_size_MISO_far}
\end{figure}
Lastly,
we consider the MISO case
where the BS is equipped with a UPA.
Fig. \ref{snr_numerical_results_IRS_size_MISO_far} plots 
the SNR versus the IRS size,
based on the summation in \eqref{eq_far_max_snr}
and closed-form bounds of the SNR 
in \eqref{eq_far_miso_bounds}.
The directional gain pattern 
of IRS's reflecting elements
is based on \eqref{eq_element_pattern_effective_aperture}.
The transmit SNR is $\bar{P}=90\,\mathrm{dB}$, 
and the center of the BS and user are located at 
$(r_q,\theta_q,\phi_q)=\left(10^3\,\mathrm{m},\frac{\pi}3,-\frac{\pi}4\right)$
and
$(r_p,\theta_p,\phi_p)=\left(10\,\mathrm{m},\frac{\pi}2,0\right)$,
respectively.
It is observed that 
the derived closed-form bounds 
can tightly estimate the SNR value 
when the IRS size is not very large.
Furthermore,
there exists a notable gap between the conventional UPW model 
and the accurate near-field model as the IRS size goes large.

\section{Conclusions}
This paper studied 
the near-field modelling and performance analysis 
for wireless communication with XL-IRS.
A generic near-field modelling for XL-IRS was developed,
by taking into account the directional gain pattern 
of IRS's reflecting elements 
and the variations in received signal amplitude 
across different reflecting elements.
We firstly derived 
tight lower- and
upper-bounds of the received SNR for the UPA-based XL-IRS.
To gain more insights,
the special case of ULA-based XL-IRS was further considered,
together with the extension of our developed near-field modelling 
to the MISO setup.
Numerical results verified our theoretical analysis and
demonstrated the necessity of proper near-field modelling for
wireless communications aided by XL-IRS.


%

\appendices

\section{Proof of Lemma \ref{lemma_boresight}}

When $\Phi_q,\Phi_p \ll \frac{r_q}{L_y}$ and 
$\Theta_q,\Theta_p \ll \frac{r_q}{L_z}$,
the double integral 
in \eqref{eq_siso_general_bounds_function} 
reduces to the following form
\begin{equation}
  \label{eq_appendix_A_integral}
  \begin{aligned}
    I=
    &\int_0^{2 \pi} \mathrm{d} \zeta 
    \int_0^R 
    \frac{r \mathrm{d} r}{\left[\left(1+\frac{r^2}{r_q^2}\right)\left(1+\frac{r^2}{r_p^2}\right)\right]^{\left(q\prime+1\right) / 2}}\\
    &=2\pi r_q^2 
    \int_0^{\frac{R}{r_q}}
    \frac{r \mathrm{d} r}{\left[\left(r^2+1\right)\left(\rho^2 r^2+1\right)\right]^{\left(q\prime+1\right) / 2}}.
  \end{aligned}
\end{equation}

By letting $r=\tan\alpha$, we can transform 
\eqref{eq_appendix_A_integral} into
\begin{equation}
  \label{eq_appendix_A_alpha}
  I=2\pi r_q^2 
  \int_0^{\arctan \frac{R}{r_q}} 
    \frac{\cos^{2q\prime}\alpha \tan\alpha \mathrm{d}\alpha}{[\rho^2+(1-\rho^2)\cos^2 \alpha]^{(q\prime+1)/2}}.
\end{equation}

With Theorem \ref{theorem_bounds},
and by substituting \eqref{eq_appendix_A_alpha}
into \eqref{eq_siso_general_bounds_function},
we can have
\begin{equation}
  \label{eq_appendix_A_f}
  \begin{aligned}
  f\left(R,q\prime\right)
  &=\left(\frac{\lambda}{4\pi}\right)^4
  \frac{\gamma\prime^2 \bar{P}}{d^4 r_q^2 r_p^2}
  I^2\\
  &=\frac{\mu^2 \bar{P}}{4 d^4}
  \left[ \rho
    \int_0^{\arctan \frac{R}{r_q}} 
    \frac{\cos^{2q\prime}\alpha \tan\alpha \mathrm{d}\alpha}{\left[\rho^2+(1-\rho^2)\cos^2 \alpha\right]^{(q\prime+1)/2}}
  \right]^2\\
  &=\frac{\mu^2 \bar{P}}{4 d^4}G(R,q\prime),
  \end{aligned}
\end{equation}
where the facts
$\gamma\prime=\frac{4\pi}{\lambda^2} \mu$ and $\rho = r_q/r_p$ are used,
and then Lemma \ref{lemma_boresight} can be obtained.

\section{Proof of Lemma \ref{bounds_square_pattern}}

By letting $q\prime=1$,
\eqref{eq_G_integral} can be expressed as 
\begin{equation}
  \label{eq_appendix_B_G}
  \begin{aligned}
    G(R,1)
    &=\left[ \rho
    \int_0^{\arctan R/r_q} 
    \frac{\cos^2 \alpha \tan\alpha \mathrm{d}\alpha}{\rho^2+(1-\rho^2)\cos^2 \alpha}
    \right]^2\\
    &=\left[ \rho
    \int_0^{\arctan R/r_q} 
    \frac{\cos \alpha \mathrm{d}(\cos\alpha)}{\rho^2+(1-\rho^2)\cos^2 \alpha}
    \right]^2.
  \end{aligned}
\end{equation}

First, for $\rho=1$, \eqref{eq_appendix_B_G} can be simplified as 
\begin{equation}
  \label{eq_appendix_B_rho_1}
  \begin{aligned}
    G(R,1)
    =\left[\int_0^{\arctan R/r_q} 
    \cos \alpha \mathrm{d}(\cos\alpha)
    \right]^2
    =\frac14 
    \Bigg[\cos^2\left(\arctan \frac{R}{r_q}\right)-1
    \Bigg]^2.
  \end{aligned}
\end{equation}

Next, for $0<\rho<1$, \eqref{eq_appendix_B_G} can be 
further expressed as
\begin{equation}
  \label{eq_appendix_B_rho}
  \begin{aligned}
    G(R,1)
    &=\left[ \frac{\rho}{1-\rho^2}
    \int_0^{\arctan R/r_q} 
    \frac{(1-\rho^2)\cos \alpha \mathrm{d}(\cos\alpha)}{\rho^2+(1-\rho^2)\cos^2 \alpha}
    \right]^2\\
    &=\frac{\rho^2}{4\left(1-\rho^2\right)^2}
    \Bigg[
    \ln \left[\rho^2+(1-\rho^2)\cos^2\left(\arctan \frac{R}{r_q}\right)\right]  
  \Bigg]^2.
  \end{aligned}
\end{equation}

Then, the proof of Lemma \ref{bounds_square_pattern} is completed.

\section{Proof of Lemma \ref{bounds_MISO_case}}

With $\Phi_p \ll \frac{r_p}{L_y}$, $\Theta_p \ll \frac{r_p}{L_z}$
and $q\prime=\frac12$,
the double integral in \eqref{eq_far_max_snr_integral_bounds}
can reduce to 
\begin{equation}
  \label{eq_appendix_C_integral}
  \begin{aligned}
    I\prime
    &= \int_o^{2\pi} \mathrm{d} \zeta \int_0^R
    \frac{r \mathrm{d} r}{\left(1+\frac{r^2}{r_p^2}\right)^{3/4}}
    =2\pi \left[\left(1+\frac{r^2}{r_p^2}\right)^{1/4}\right]_0^R\\
    &=2\pi \left(\sqrt[4]{\frac{R^2}{r_p^2}+1}-1\right).
  \end{aligned}
\end{equation}

By substituting \eqref{eq_appendix_C_integral} 
into \eqref{eq_far_max_snr_integral_bounds},
we obtain
\begin{equation}
  \label{eq_appendix_C_U}
  \begin{aligned}
  U\left(R,\frac12\right)
  =N\left(\frac{\lambda}{4\pi}\right)^4
  \frac{\gamma\prime^2 \Psi_q\bar{P}}{d^4 r_q^2 r_p^2}
  I\prime^2
  =N\bar{P}
  \frac{\lambda^4 \gamma\prime^2 \Psi_q r_p^2}{16 \pi^2 d^4 r_q^2}
  \bigg(\sqrt[4]{\frac{R^2}{r_p^2}+1}-1\bigg)^2.
  \end{aligned}
\end{equation}

Thus, the proof of Lemma \ref{bounds_MISO_case} is completed.




\ifCLASSOPTIONcaptionsoff
  \newpage
\fi

\bibliographystyle{IEEEtran}
\bibliography{ref}

\end{document}